\documentclass[nofootinbib,twocolumn,superscriptaddress,preprintnumbers]{revtex4-1}
\usepackage[utf8]{inputenc}
\usepackage{amsfonts}
\usepackage{mathrsfs}
\usepackage{physics}
\usepackage{footnote}
\usepackage{scrextend}
\usepackage{leftidx}
\usepackage{soul}
\usepackage{makecell}
\usepackage{tabularx}
\usepackage{cprotect}

\usepackage[dvipsnames]{xcolor}
\definecolor{red}{rgb}{0.9, 0,0}
\definecolor{cerulean}{rgb}{0., 0.42,0.9}
\definecolor{prettygreen}{rgb}{0., 0.55,0.3}

\usepackage{epsfig}
\usepackage{graphicx}  
\usepackage{url}
\usepackage{color}
\usepackage{hyperref}
\hypersetup{
     colorlinks   = true,
     citecolor    = red,
	 linkcolor=blue
}
\usepackage{float}
\usepackage{placeins}

\newcommand{\bfq}{{\bf q}}
\newcommand{\bfp}{{\bf p}}

\newcommand{\bfk}{{\bf k}}

\newcommand{\bfK}{ {\bf K} }
\newcommand{\bfv}{{\bf v}}

\newcommand{\GeV}{\mathrm{GeV}}
\newcommand{\MeV}{\mathrm{MeV}}

\newcommand{\Zion}{ Z_{\rm ion} }

\def\beq{\begin{eqnarray}}
\def\eeq{\end{eqnarray}}
\def\bea{\begin{eqnarray}}
\def\eea{\end{eqnarray}}

\definecolor{red}{rgb}{0.9, 0,0}
\definecolor{cerulean}{rgb}{0., 0.5,0.8}

\newcommand{\code}{\emph{DarkELF}}

\graphicspath{{./figures/}}

\begin{document}

\title{
DarkELF: A python package for dark matter scattering in dielectric targets
}

\author{Simon Knapen}
\email{simon.knapen@cern.ch}
\affiliation{CERN, Theoretical Physics Department, Geneva, Switzerland}
\author{Jonathan Kozaczuk}
\email{jkozaczuk@physics.ucsd.edu}
\affiliation{Department of Physics, University of California, San Diego, CA 92093, USA}
\author{Tongyan Lin}
\email{tongyan@physics.ucsd.edu}
\affiliation{Department of Physics, University of California, San Diego, CA 92093, USA}

\date{\today}

\begin{abstract}
We present a python package to calculate interaction rates of light dark matter in dielectric materials, including screening effects. The full response of the material is parametrized in the terms of the energy loss function (ELF) of material, which \code\ converts into differential scattering rates 
for both direct dark matter electron scattering and through the Migdal effect. In addition, \code\ can calculate the rate to produce phonons from sub-MeV dark matter scattering via the dark photon mediator, as well as the absorption rate for dark matter comprised of dark photons. The package includes precomputed ELFs for Al, $\mathrm{Al}_2\mathrm{O}_3$, GaAs, GaN, Ge, Si, $\mathrm{SiO}_2$, and ZnS, and allows the user to easily add their own ELF extractions for arbitrary materials.
\end{abstract}

\maketitle

\tableofcontents
\section{Introduction}

The search for the direct detection of the dark matter has progressed to a phase where there are numerous experiments aiming to probe sub-GeV dark matter (DM), often by leveraging electronic excitations, see e.g.~\cite{Barak:2020fql,Castello-Mor:2020jhd,Amaral:2020ryn,Bernstein:2020cpc}. In addition the next generation of detectors is aiming for energy thresholds well below the ionization threshold of the target \cite{Hertel:2018aal,Fink:2020noh}, thus opening the path to search for individual phonon excitations.  
For all such strategies, the many-body physics of the target material is important and detailed calculations at the interface with condensed matter physics are therefore needed to accurately extract the relevant scattering rates. 

Electron excitations may arise from direct DM-electron scattering~\cite{Essig:2011nj,Essig:2012yx,Essig:2015cda}, as shake-off electrons from nuclear recoils~\cite{Vergados:2004bm,Moustakidis:2005gx,Ejiri:2005aj} or from secondary ionizations as the recoiling nucleus travels through the target material. Solid state targets are particularly advantageous because they can have arbitrarily small gaps to produce electron excitations. However, because their electron wavefunctions are delocalized and highly non-trivial, calculations of the differential scattering rate are often involved and material dependent. For Si and Ge targets, Essig~et al.~\cite{Essig:2015cda}   performed the first calculation of DM-electron scattering using electronic wavefunctions obtained with density functional theory (DFT). This calculation was subsequently applied to a broader range of semiconductors~\cite{Griffin:2019mvc,Trickle:2019nya}. 

It was recently pointed out that the DM-electron scattering rate can be extracted directly from the energy loss function (ELF) 
\begin{align}
    \Im \left[ \frac{-1}{\epsilon(\omega,\bfk)} \right]
\end{align}
of the target material~\cite{Knapen:2021run,Hochberg:2021pkt}, where $\epsilon(\omega,\bfk)$ is the momentum and frequency dependent longitudinal dielectric function. This approach has two main advantages: \emph{(i)} In-medium screening effects are automatically included and were found to reduce the scattering rate by a non-negligible amount~\cite{Knapen:2021run}
\emph{(ii)} The ELF is exceptionally well-studied experimentally and theoretically in the materials science literature, which means that standard and well validated tools can be used to extract it for the target of interest. In \cite{Knapen:2021run}, we calculated the ELF for Si and Ge using time-dependent density functional theory (TDDFT) methods with the GPAW package~\cite{GPAW1, GPAW2} and compared this method with an approach fitting data to a Mermin oscillator model  \cite{PhysRevB.1.2362,chapidif}. We elaborate on these methods and their advantages and shortcomings in Sec.~\ref{sec:ELF}. We found both methods to be in excellent agreement in the regime most relevant for DM-electron scattering, as discussed in Sec.~\ref{sec:DMelectron}.
 
Even if the DM couples predominantly to nuclei, it can still leave an electronic signal in the detector. One way this could happen is if the nucleus ``shakes-off'' an electron during the initial hard recoil \cite{Vergados:2004bm,Moustakidis:2005gx,Ejiri:2005aj}. This is known as the Migdal effect \cite{Migdal1939,Migdal:1977bq}, and has been applied extensively to DM scattering off atomic targets~\cite{Bernabei:2007jz,Ibe:2017yqa,Dolan:2017xbu,Essig:2019xkx,Bell:2019egg,Liang:2019nnx,Baxter:2019pnz,GrillidiCortona:2020owp,Nakamura:2020kex,Liu:2020pat}.  Refs.~\cite{Knapen:2020aky,Liang:2020ryg} provided the first full derivation of the Migdal effect for dark matter scattering in semiconductors, showing that it can be treated as an in-medium analog of bremsstrahlung. The ELF again plays a critical role, as it determines the probability for the nucleus to shake-off an electron. In Sec.~\ref{sec:DMMigdal} we summarize our results of \cite{Knapen:2020aky}, present a few new computations, and discuss their implementation in \code. 

Neither the Migdal effect nor direct electron recoils are available if the kinetic energy of the DM is below the electron bandgap of the target. In this case the dark matter can still deposit energy by producing one or more athermal phonons in the target. Such processes have been studied extensively in both superfluid He \cite{Schutz:2016tid,Knapen:2016cue,Acanfora:2019con,Caputo:2019cyg,Caputo:2019xum,Caputo:2020sys} and solid state targets \cite{Knapen:2017ekk,Griffin:2018bjn,Campbell-Deem:2019hdx,Cox:2019cod,Trickle:2019nya,Griffin:2019mvc,Mitridate:2020kly,Trickle:2020oki,Coskuner:2021qxo}. Given the existing constraints on models of sub-MeV dark matter, DM scattering through a dark photon mediator and dark photon DM absorption appear to be the most promising processes \cite{Knapen:2017xzo}. Both are most pronounced in polar materials \cite{Knapen:2017ekk,Griffin:2018bjn} and can be modeled with the ELF, for frequencies below the band gap of the target. Previous calculations rather heavily relied on computationally intensive DFT methods, though analytical approximations are available some instances. Here we present an intermediate method, where we write the rate in terms of the ELF, which we subsequently take from experimental data. For phonon-scattering and absorption processes we moreover only need to know the ELF in the low momentum (optical) limit, for which good experimental measurements are readily available. The ELF method is more accurate than the existing analytical approximations, while bypassing the time-consuming DFT calculations. DM-phonon scattering and dark photon absorption are discussed in Sec.~\ref{sec:DMphonon} and Sec.~\ref{sec:absorption} respectively.

\code\ is available at
\begin{center}
\url{https://github.com/tongylin/DarkELF}
\end{center}
and comes with tabulated ELFs for $\mathrm{Al}_2\mathrm{O}_3$, GaN, Al, ZnS, GaAs, $\mathrm{SiO}_2$, Si and Ge, allowing the user to easily calculate differential DM scattering rates in these materials. Additional materials and ELF extractions may be added to the repository as the need arises. Users can compute the rate subject to various fiducial cuts, or implement their own form factors to study non-standard DM models. 
It is also straightforward for a user to add their own calculations or extractions of the ELF, facilitating fast comparisons between methods and materials. This makes \code\ also a suitable tool for target optimization and to study the theoretical uncertainties associated with the scattering rate. In Sec.~\ref{sec:conclusions} we offer some concluding remarks and comment on possible future additions to the code. For instruction on the usage of \code, we refer to Apprendix~\ref{app:useage} and the example \texttt{jupyter} \cite{jupyter} notebooks in the repository. The example notebooks also contain a number of additional plots which were omitted in the paper for brevity.

\section{Calculating the ELF\label{sec:ELF}}

\begin{figure*}
\includegraphics[width=\textwidth]{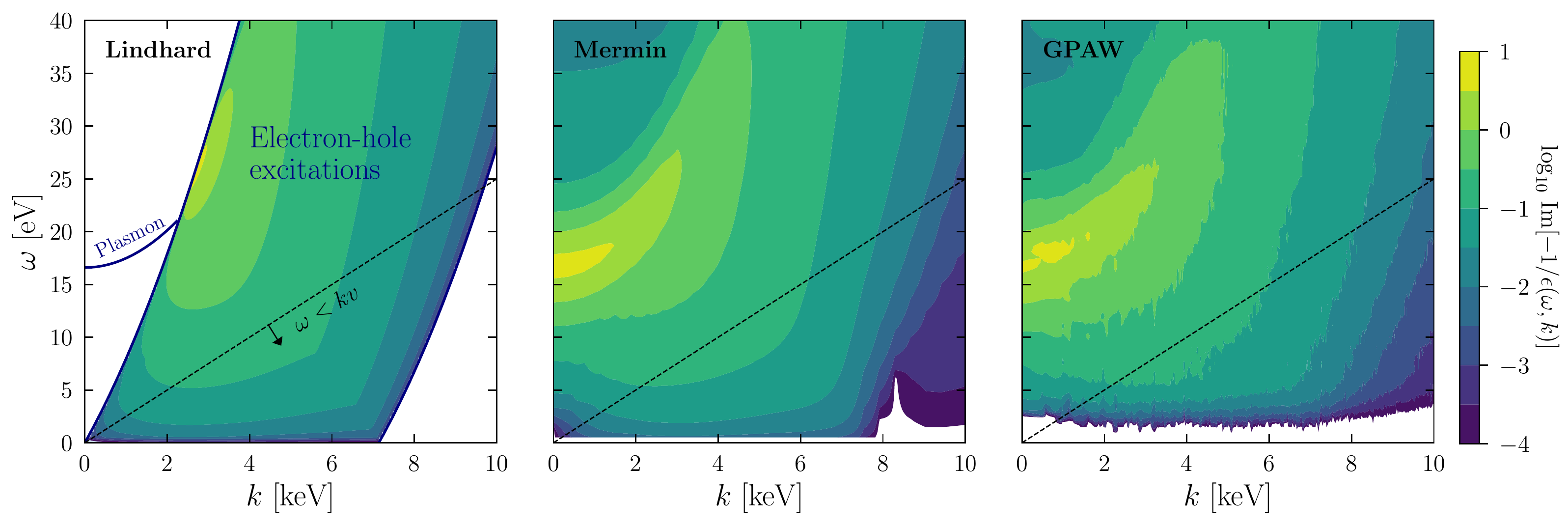}
\caption{ELF for Si, calculated using the Lindhard, Mermin and GPAW methods, as described in the text. The blue line in the left-hand panel indicates the location of plasmon pole, which is a Dirac delta-function in the Lindhard method. Only the GPAW method (right-hand panel) correctly models the low $\omega$ regime, close to the band gap. For halo DM scattering off electrons, the accessible phase space is bounded by $\omega < k v$, which is indicated by the dashed line with $v = 2.5 \times 10^{-3}$. \label{fig:elf_full}}
\end{figure*}

The ELF describes the energy loss of a charged particle traveling through the material. It is therefore not only of practical importance, but also provides a window into the physical mechanisms at play in the target. Furthermore detailed first-principles calculations of the ELF are now possible, which can be compared with experimental data. For our purposes, this means that there a number of complementary methods which one can use to compute the ELF, and comparing them can give us some insight in the uncertainties associated with dark matter interactions with the target material. 

\code~is set up independently of the method used to calculate the ELF, as the real and imaginary components of the dielectric function ($\epsilon_1,\epsilon_2$) are read in as a look-up table.  The user can therefore supply their own calculation of the ELF and straightforwardly extract the dark matter interaction rates. We also supply a number of precomputed  look-up tables with the code. Note that everywhere in this work, dielectric function refers exclusively to the longitudinal dielectric function. We will also work in the approximation that both the ELF and dielectric function are isotropic in momentum and diagonal in reciprocal lattice space~\cite{Knapen:2021run}.

In this section, we focus on the ELF for $\omega >  E_{\rm gap}$, with $E_{\rm gap}$ the electron band gap, where the energy loss is dominated by the electron response of the material. Below the electron band gap, the leading contribution to the ELF will generally be phonons, which will be discussed in Sec.~\ref{sec:DMphonon}. For the electron-response regime, we supply results for three independent methods to compute the ELF:
\begin{itemize}
\item The \textbf{Lindhard method} is the most simplistic and uses the Lindhard dielectric function, which models the material as a non-interacting Fermi liquid. The main advantage of using the Lindhard dielectric function is its simplicity, as it depends only on the plasma frequency   ($\omega_p$), or equivalently, the Fermi velocity \cite{DresselGruner}:
\begin{align}\label{eq:eps_lindhard_gas}
    \epsilon_{\rm Lin}(\omega,k)=&1+\frac{3\omega_p^2}{k^2 v_F^2}\lim_{\eta\to0}\Bigg[f\left(\frac{\omega+i\eta}{k v_F},\frac{k}{2 m_e v_F}\right)\Bigg]
 \end{align}
with
\begin{align}
v_F&=\left(\frac{3\pi \omega_p^2}{4 \alpha m_e^2}\right)^{1/3}\nonumber\\
f(u,z)&=\frac{1}{2}+\frac{1}{8z}\left[g(z-u)+g(z+u)\right]\nonumber\\
g(x)&=(1-x^2)\log\left(\frac{1+x}{1-x}\right)\nonumber
\end{align}
with $\alpha$ and $m_e$ respectively the fine structure constant and the electron mass. The Lindhard dielectric function approximates the material as homogeneous and neglects all dissipation effects.
This means that the plasmon pole is infinitely narrow, an approximation which is badly violated in most semiconductors. For halo DM, however, scattering is dominated by the production of electron-hole pairs far away from the plasmon pole, which can be modeled qualitatively with the Lindhard ELF. This is shown in the left-hand panel of Fig.~\ref{fig:elf_full}. The Lindhard ELF does not provide an accurate description of realistic semiconductors at low $k$ and high $\omega$, and therefore cannot be used for absorption processes.

\item The \textbf{Mermin method} is a generalization of the Lindhard method which includes dissipation and can also be used for absorption processes. 
Concretely, a dissipation parameter $\Gamma$ can be added to the Lindhard model in a self-consistent way by defining the Mermin dielectric function \cite{PhysRevB.1.2362}  
\begin{equation}\label{eq:mermin}
    \epsilon_{\text{Mer}}(\omega,k) = 1+\frac{(1+i \frac{\Gamma}{\omega})(\epsilon_{\text{Lin}}(\omega+i\Gamma,k)-1)}{1+(i\frac{\Gamma}{\omega})\frac{\epsilon_{\text{Lin}}(\omega+i\Gamma,k)-1}{\epsilon_{\text{Lin}}(0,k)-1}}.
\end{equation}
In the Mermin method, the ELF is modeled as a superposition  of ELFs obtained with the Mermin dielectric function, where the plasma frequencies, dissipation parameters and the weights of the different terms are  fitted to experimental data. In an ad hoc way, this weighted linear combination accounts for the inhomogeneities in the electron number density within the unit cell. The fitted data typically includes the  measured ELF from reflection electron energy loss spectroscopy (REELS) and/or optical data \mbox{($k=0$ limit)}, and therefore can reproduce absorption processes. The theoretically motivated ansatz in \eqref{eq:mermin} provides a way to perform a controlled extrapolation of the ELF to finite $k$, while conserving local electron number. Experimental collaborations \cite{NOVAK20087,Dingdatabase,TU2007865} moreover occasionally present their results in terms of fits to models whose parameters can be reinterpreted in terms of the Mermin model.  This reinterpretation is done with the \verb+chapidif+ package \cite{chapidif}, which builds on the work in \cite{10.1002/sia.6227,PhysRevA.58.357,VOS2019242}.  For more details about our procedure we refer to our earlier work in \cite{Knapen:2021run}. 

The middle panel of Fig.~\ref{fig:elf_full} shows the ELF for Si, as obtained with the Mermin method applied to the experimental data in \cite{Dingdatabase}. The low $k$ region near the plasmon pole is much more realistic than for the Lindhard ELF, as this is the regime where the ansatz is fit to the experimental data. Even with a finite width, the plasmon region is still well outside the kinematically allowed regime for DM-electron scattering, as indicated by the dashed black line. The Mermin method however does not incorporate the detailed band structure of the material. In particular, in the middle panel of Fig.~\ref{fig:elf_full} one can see that it effectively predicts a vanishing band gap, which is of course not realistic for a semiconductor such as Si.\footnote{The band gap can be approximated by the ad hoc addition of a Heaviside step function $\theta(\omega - E_{\text{gap}})$  \cite{ABRIL20101763} or with the Mermin-Levine-Louie ansatz (MLL) \cite{PhysRevB.25.6310}. See \cite{VOS2017192,10.1002/sia.6227} for comparisons between these various approaches. }As we will see, it is also less appropriate to model the high momentum ($k\gtrsim 15$ keV) regime.

\begin{figure*}
\includegraphics[width=\textwidth]{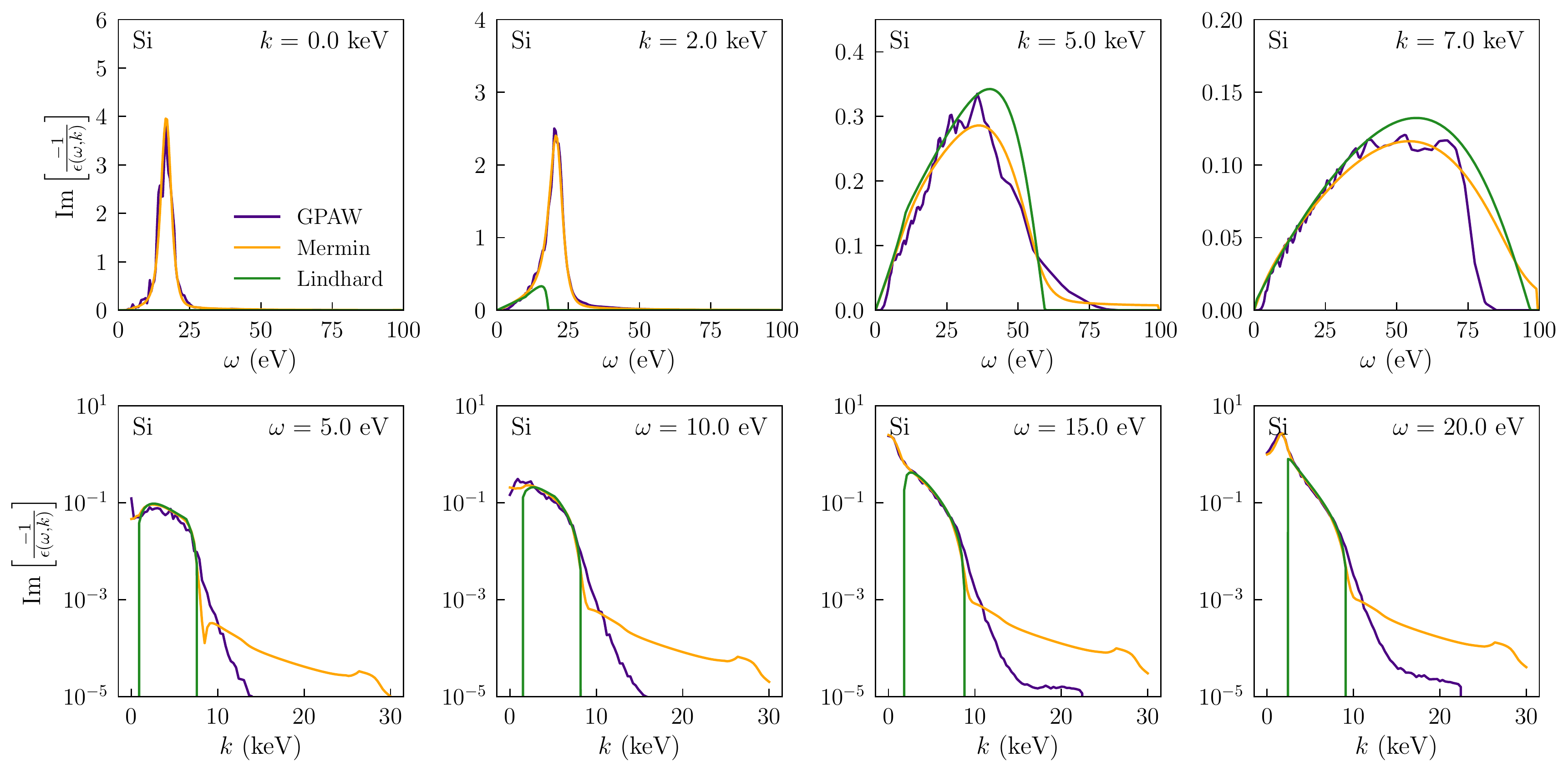}
\caption{ELF for Si, calculated using the Lindhard, Mermin and GPAW methods for select values of $\omega$ and $k$. \label{fig:ELFsliced} }
\end{figure*}

\item The \textbf{GPAW method} is the most sophisticated of the three methods we employ, as it relies on a first-principles TDDFT calculation with the software package \verb+GPAW+~\cite{GPAW1, GPAW2}. In this method one approximates the many-body electron wave functions with a  Kohn-Sham (KS) system~\cite{PhysRev.140.A1133} of effective, single particle wave functions subject to an effective potential. This system is then solved numerically on a periodic lattice. The GPAW method does the best job in modeling the detailed properties of the material, in particular for $\omega$ near the band gap. This is shown in the right-hand panel of Fig.~\ref{fig:elf_full}, where the band gap is now clearly visible at low $\omega$. The GPAW method is however by far the most computationally intensive of the three and is most the difficult to validate for non-experts in TDDFT methods. At this time we therefore only provide ELF look-up tables calculated with the GPAW method for Si and Ge.\footnote{At this time, \code\ only accounts for the diagonal part of the ELF, which more generally is a matrix in reciprocal lattice space. Throughout this paper, $\bfk$ therefore indicates an arbitrary momentum vector which can be outside the first Brillouin zone, such that $\bfk\equiv\bfk'+\bfK$ with $\bfk'$ restricted to the first BZ and $\bfK$ a reciprocal lattice vector. See \cite{Knapen:2021run} for details.} For more details on our calculations of the ELF in GPAW, we refer the reader to \cite{Knapen:2021run}.
\end{itemize}

The limitations and regime of applicability of each method can be made more manifest by taking slices for fixed $k$ and $\omega$, as shown in Fig.~\ref{fig:ELFsliced}.  In the low $k$ regime (upper left panels) the plasmon peak is clearly visible and we find excellent agreement between the Mermin and GPAW methods. The Lindhard method on the other hand fails spectacularly due to its omission of dissipation effects. For higher values of $k$ (upper right panels) we are firmly in the electron-hole pair regime and all three methods are in fairly good agreement for $\omega\lesssim 25$ eV. The Lindhard method remains in qualitative agreement with the others for higher $\omega$ as well, though the approximation is clearly less suitable. Our calculations with the GPAW method are not applicable beyond $\omega\gtrsim 75$ eV since only the 70 lowest laying bands were included for computational reasons.  

In the left-hand panel of Fig.~\ref{fig:elf_full} and the bottom row of Fig.~\ref{fig:ELFsliced} we see the well-known fact that the Lindhard model does not allow for excitations of electron-hole pairs to be created outside a band in momentum space, whose width is set by the Fermi momentum. We will refer to this band as the Lindhard electron-hole continuum. In the bottom row of Fig.~\ref{fig:ELFsliced} we see that all three methods are in good agreement within this region. In the Mermin and GPAW methods, excitations outside the Lindhard electron-hole continuum are  allowed.  At lower $k$ values, the Mermin and GPAW methods are also in reasonably good agreement with each other provided that $\omega\gtrsim 5$ eV, well above the electron band gap. For $k$-values above the Lindhard electron-hole continuum ($k\gtrsim10$ keV) the Mermin and GPAW methods start diverging rather strongly. Both methods are challenged here: For the GPAW method one needs an increasingly large grid in momentum space, which significantly impacts the computational requirements of the calculation. In our calculations we restricted the grid to $k\lesssim 22$ keV, which corresponds to the sharp edge in the two bottom right panels of Fig.~\ref{fig:ELFsliced}. Beyond this value we currently do not make a prediction for the ELF, and \code\ will automatically restrict the phase space of all processes to $k$ values satisfying this constraint. 

The Mermin method reproduces the measured Compton spectrum for high momenta ($k\gtrsim 20$ keV) and high energy ($\omega\gtrsim 1$ keV) \cite{VOS20166},
though its validity for high $k$ and low $\omega$ regime that is of interest for dark matter scattering is less established. In particular, the lower row of Fig.~\ref{fig:ELFsliced} shows that the Mermin method predicts a substantially larger ELF in the high $k$ regime than the GPAW method. As we will see in the next section, this regime is relevant for dark matter experiments with energy thresholds exceeding roughly 15 eV. This behavior as predicted by the Mermin method is likely not accurate and can be traced back to the rather rigid functional form in \eqref{eq:mermin}, as both the $k$ regime above and below the Lindhard electron-hole continuum are controlled by the same set of dissipation parameters. Moreover, the various fits only take into account optical and/or REELS data, which are both probing the low momentum regime.\footnote{REELS measurements do give access to finite $k$, see e.g.~\cite{PhysRevB.100.245209}, but the rate is still dominated by low to intermediate $k$. The unfolding of the experimental results with the inverse Monte Carlo method in \cite{PhysRevB.100.245209} therefore likely leads to large systematic uncertainties in the high $k$ regime. Finally, their results are not publicly available in a suitable format and are thus currently not included in our analysis.}

Inelastic X-ray scattering measurements on the other hand are a good alternative in the high $k$ regime. Weissker et.~al.~\cite{Weissker} carried out a series of such ELF measurements for Si at European Synchrotron Radiation Facility (ESRF) with an 8 keV X-ray beam. Unfortunately, the momentum transfer they had access to is insufficiently high to diagnose the discrepancy in Fig.~\ref{fig:ELFsliced}. In the regime which they do have access to, our calculations are in good agreement with their measurements, see \cite{Knapen:2021run}. They moreover carefully compare their results with a suite of TDDFT calculations and find overall good agreement as well. For these reasons, we assign more credence to our GPAW result in the high $k$ regime, but independent experimental verification with data from a high energy synchotron facility would nevertheless be interesting. In the next sections we will comment in some detail on how these various uncertainties propagate to the dark matter scattering rate.

\section{Dark matter-electron scattering \label{sec:DMelectron} }

In the section, we briefly summarize the formalism for dark matter-electron scattering as laid out in \cite{Knapen:2021run}. We illustrate the functionality and limitations of calculations with \code\ by comparing results obtained with the Lindhard, Mermin and GPAW methods, as well as a number of different materials. We do not, however, attempt an exhaustive comparison between possible target materials in this paper.

\begin{figure*}[t]
\includegraphics[width=\textwidth]{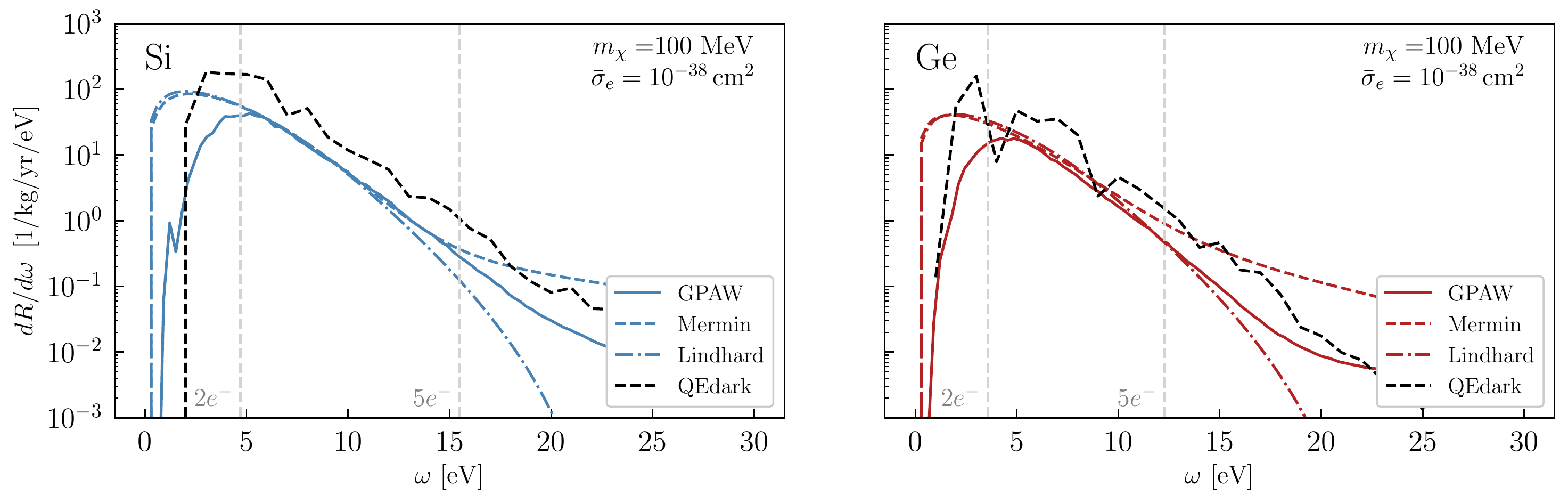}
\caption{Comparison between Lindhard, Mermin and GPAW calculations of the differential scattering rate for Si and Ge, for the massless mediator regime. The vertical dashed lines indicate the 2$e^-$ and 5$e^-$ thresholds. The black dashed line is obtained with the \texttt{QEdark} code \cite{Essig:2015cda}, which does not include screening effects and gives a larger rate. \label{fig:erecoilgpawvsmermin}}
\end{figure*}

To start, we assume that in the non-relativistic limit, DM of mass $m_\chi$ interacts dominantly with the electron number density $n$ by means of a mediator particle $\phi$. The interaction Hamiltonian is then
\begin{align}\label{eq:interaction}
  H=   \ g_\chi \phi \bar\chi\chi + g_e \phi n 
\end{align}
with $n$ the electron number density operator. The mediator $\phi$ could represent a scalar mediator or the time-like component of a vector mediator such as a dark photon. For this class of models, the dark matter scattering can be written in terms of the the dynamic structure factor $S(\omega,\bfk)$, which is defined as
\begin{align}\label{eq:skodef}
     S(\omega,\bfk) \equiv
    \frac{2 \pi}{V} \sum_{i,f} P_i| \langle f &| n_{-\bfk} | i \rangle |^2 \delta(\omega + E_i - E_f). 
\end{align}
with $n_{-\bfk}$ the Fourier transform of the electron number density operator
and $P_i\equiv e^{-\beta E_i}/Z$ is the thermal occupation number. Here $\beta$ is the inverse temperature ($\beta\equiv 1/T$), $Z$ the partition function of the system and $V$ its volume. The initial and final states of the system are denoted by $| i \rangle$ and $\langle f |$ respectively, with corresponding energies $E_i$ and $E_f$.
By making use of the fluctuation-dissipation theorem, one can show that the dynamical structure function is related to the ELF by \cite{PhysRev.113.1254}
\begin{align}
S(\omega,\bfk)=    \frac{k^2}{2\pi \alpha} \frac{1}{1 - e^{-\beta \omega}}  \Im \left[ \frac{-1}{\epsilon(\omega,\bfk)} \right].
   \label{eq:ELF_structurefactor}
\end{align}
Also folding in the DM velocity distribution, DM scattering form factor and the various flux factors, we arrive at our final expression for the DM scattering rate, in units of number of counts per unit of exposure
\begin{widetext}
\begin{align}\label{eq:mastereq}
	R = \frac{1}{\rho_T} \frac{\rho_\chi}{m_\chi} \frac{  \bar \sigma_e}{\mu_{\chi e}^2} \frac{\pi}{\alpha} \int\! d^3 v\, f_\chi(v)   \int\!\! \frac{ d^3 \bfk}{(2\pi)^3} k^2   |F_{DM}(k)|^2 \int\!\frac{d\omega}{2\pi}\,  \,   \frac{1}{1 - e^{-\beta \omega}} \Im \left[ \frac{-1}{\epsilon(\omega,\bfk)} \right]\delta\left( \omega + \frac{k^2}{2 m_\chi} - \bfk \cdot \bfv \right).
\end{align}
\end{widetext}
where $f_\chi(v)$ is the dark matter velocity distribution, which is taken to correspond to the Standard Halo Model with $v_{\rm esc} = 500$ km/s, velocity dispersion $v_0 = 220$ km/s, and Earth velocity $v_e = 240$ km/s. $\rho_\chi$ is the local DM density, taken to be 0.4 GeV/cm$^3$. The DM-mediator form factor is defined as
\begin{align}\label{eq:DMformfactor}
    F_{DM}(k) = \frac{  \alpha^2 m_e^2+m_{\phi}^2}{ k^2+m_{\phi}^2 }.
\end{align}
The limiting cases of \mbox{$F_{DM}(k)=1$} and \mbox{$F_{DM}(k)=\alpha^2 m_e^2/k^2$} are most frequently studied and are referred to respectively as the ``massive mediator'' and ``massless mediator'' regimes. The user moreover has access to the double differential distribution $d^2 R/dkd\omega$, such that more general form factors can be implemented easily. The effective cross section is defined as
\begin{align}\label{eq:sigmaedefinition}
	\bar \sigma_e = \frac{\mu_{\chi e}^2 g_e^2 g_\chi^2}{\pi \big(\alpha^2 m_e^2 + m_{\phi}^2\big)^2}.
\end{align}

In general, the ELF can depend on the direction of the momentum transfer $\bfk$, though for many materials the isotropic approximation is very good. The current version of \code\, therefore assumes the isotropic limit for $S(\omega,\bfk)$ and the ELF. The generalization to the non-isotropic case is left for future work. The functions provided by \code\, are summarized in Appendix~\ref{app:useage}. In the functions which compute the (differential) rate, the velocity integral has already been performed, by swapping the order of the integrals in \eqref{eq:mastereq}. 

\begin{figure*}[t]
\includegraphics[width=\textwidth]{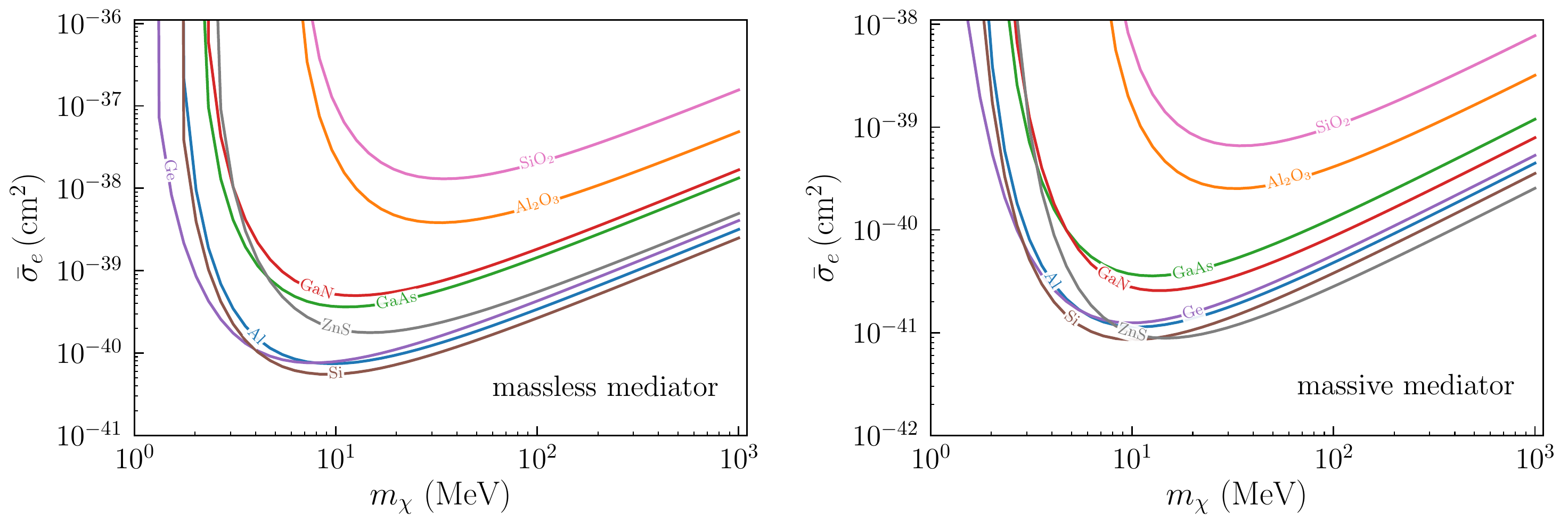}
\caption{For DM-electron scattering, a comparison of the cross sections needed to obtain 3 events for a kg-year exposure. For all for 8 materials, the ELF is obtained with the Mermin method. The threshold was taken to be the 2$e^-$ threshold for Ge, Si and GaAs, 5 eV for Al and $2\times E_{\mathrm{gap}}$ for the remaining materials. For the massive mediator, we restricted the phase space to $k<12$ keV. As such, these cross sections curves should be viewed as a conservative upper bound. \label{fig:erecoilmatcomp}}
\end{figure*}

Fig.~\ref{fig:erecoilgpawvsmermin} shows the comparison of the differential scattering rate obtained with the Lindhard, Mermin and GPAW methods for a benchmark model point. As explained in the previous section, the Lindhard and Mermin methods are less reliable for $\omega$ values near the band gap ($E_{\text{gap}}$), which results in the disagreement for $\omega\lesssim$ 5 eV. If we follow the treatment in \cite{Essig:2015cda,doi:10.1063/1.1656484} to convert $\omega$ into the observed number of ionization electrons, this roughly corresponds to the $2e^-$ threshold. Hence, if the single ionization electron rate is desired, the GPAW method is recommended. That said, even with the GPAW method our current ELF grids are fairly noisy for $\omega\lesssim 2 E_{\text{gap}}$, and we expect there to be $\mathcal{O}(1)$ theoretical uncertainties. 
In most experiments, however,  large backgrounds are expected in the single electron bin, and the bulk of the sensitivity will come from events with at least two ionization electrons. With a $2e^-$ threshold and a massless mediator, we find that the Mermin and Lindhard calculations are in good agreement with the GPAW computations. 

The various methods start diverging for $\omega\gtrsim 15$ eV, which roughly corresponds to 5 ionization electrons in Si. This behavior is caused by the discrepancy at high momenta in the bottom row of Fig.~\ref{fig:ELFsliced}. We recommend the GPAW results in this regime.
Note that this part of the spectrum is very subleading and is only relevant for experiments with a relatively high energy threshold, or in a post-discovery scenario where one would want to infer DM properties from the shape of the recoil spectrum. Finally, the dashed black curve on Fig.~\ref{fig:erecoilgpawvsmermin} indicates the prediction using the \verb+QEdark+ code \cite{Essig:2015cda}. The discrepancy is due to screening effects which are neglected in the \verb+QEdark+ calculation. We refer to \cite{Knapen:2021run} for a more detailed discussion of this effect. For a massive mediator, the agreement between the three methods is less satisfactory, since the rate is weighted more towards the high $k$ part of the phase space.  We thus recommend to use the ELF obtained with GPAW for the massive mediator. Plots of the differential rate for this scenario can be found on our github repository.

We compare the overall fiducial cross section reach of a number of materials in Fig.~\ref{fig:erecoilmatcomp}. We hereby assumed the $2e^-$ threshold for Ge, Si and GaAs and set the threshold to twice the band gap for all other materials except for Al. In the latter case we assumed 5 eV. At present we do not have GPAW results for materials other than Ge and Si, and we therefore used the Mermin method for all materials. The experimental inputs for Si, Al, $\mathrm{Al}_2\mathrm{O}_3$, ZnS and $\mathrm{SiO}_2$ were taken from the Ding~et.~al.~database \cite{Dingdatabase}. For Ge we used the Novak~et.~al.~data \cite{NOVAK20087} and the rates for GaAs and GaN were extracted from the measurements by Tung~et.~al.~\cite{TU2007865}. 

Commonly-used targets such as Ge and Si perform favorably as compared to the other materials considered here. While screening effects are stronger in lower-gap semiconductors such as Ge and Si, this is more than compensated for the lower threshold. The weaker reach for the other semiconductors is due in part to the higher $2e^-$ threshold. For instance, the average energy needed per ionization electron is much higher in GaAs compared to the otherwise similar material Ge \cite{doi:10.1063/1.1656484}, such that the $2e^-$ threshold is around 6.1 eV as compared to 3.6 eV for Ge. For the other semiconductors, the bandgaps are also higher than in Si and Ge. 

The results in Fig.~\ref{fig:erecoilmatcomp} were all obtained with the Mermin method. As noted above, in  the massive mediator limit, the rate is sensitive to the high $k$ regime in Fig.~\ref{fig:ELFsliced}, for which the reliability of the Mermin method is doubtful. For this reason, we chose to restrict the phase space by imposing $k\lesssim 12$ keV in the massive mediator plot in Fig.~\ref{fig:erecoilmatcomp}. The cross section curves shown should thus be interpreted as conservative upper bounds. To remedy this problem, a dedicated DFT calculation would be desirable for all materials of interest to the experimental community, similar to the target comparison performed by Griffin~et~al.~\cite{Griffin:2019mvc}. The latter results however do not yet include the $\mathcal{O}(1)$ screening effects. We leave such computations for future work.

Details on the usage of \code\ for electron recoil processes can be found in Appendix~\ref{app:useage}. 

\section{Nuclear recoils through the Migdal effect \label{sec:DMMigdal}}

The first generations of direct detection experiments were optimized to discover elastic nuclear recoils in a large target volume. For $m_\chi\lesssim 1$ GeV, the energy deposited in the nuclear recoil can however easily be below the detector threshold, and one either has to consider a dedicated, ultra-low threshold detector with a low mass target such as liquid He \cite{Guo:2013dt,Liu:2016qwd,Hertel:2018aal}, or make use of inelastic processes such as bremsstrahlung \cite{Kouvaris:2016afs} or the Migdal effect \cite{Vergados:2004bm,Bernabei:2007jz}. The Migdal effect \cite{Migdal1939,Migdal:1977bq} refers to the process where the atom shakes off one or more electrons immediately after being struck by an external probe, which in our case is the DM. This process was studied extensively in the context of DM scattering off atomic targets \cite{Vergados:2004bm,Moustakidis:2005gx,Ejiri:2005aj,Bernabei:2007jz,Ibe:2017yqa,Dolan:2017xbu,Essig:2019xkx,Bell:2019egg,Liang:2019nnx,Baxter:2019pnz,GrillidiCortona:2020owp,Nakamura:2020kex,Liu:2020pat} and estimates were obtained for semiconductor targets \cite{Essig:2019xkx,Liu:2020pat}.

In atomic targets, the calculation can be performed most conveniently by boosting to the rest frame of the recoiling atom and writing the matrix element in terms of the transition dipole moments for the atom. Ibe.~et.~al.~\cite{Ibe:2017yqa} comprehensively review this formalism in the context of DM scattering and numerically calculated the relevant matrix elements with the \verb+Flexible Atomic Code (FAC)+ \cite{doi:10.1139/p07-197}. Whenever we discuss the atomic Migdal effect in this work, we will be referring to the Ibe~et.~al.~computation, though others are available as well, as referenced above. \code\ incorporates the numerical form factors obtained in~\cite{Ibe:2017yqa} and can therefore be used to perform atomic Migdal calculations for select materials.  

The Migdal effect in semiconductors is more subtle, due to the delocalized nature of the electron clouds. This prevents one from using the boosting method, as the rest frame of the lattice is now a preferential frame. A full calculation in the rest frame of the lattice was completed simultaneously by us \cite{Knapen:2020aky}  and Liang~et.~al.~\cite{Liang:2020ryg} and revealed a qualitatively different answer from directly applying the Ibe~et.~al.~method to a crystal. In this work we also showed that plasmon production~\cite{Kurinsky:2020dpb,Kozaczuk:2020uzb} is included in the Migdal rate, but is very subleading for a DM candidate with a standard velocity profile.

Here we only present the final result and discuss its regime of validity and implementation in \code; for the full derivation and discussion, see \cite{Knapen:2020aky}. For a monatomic material, we found that the rate in number of counts per unit exposure is given by 
\begin{widetext}
\begin{align}\label{eq:migdalmaster}
R=&\frac{8\pi^2 \Zion^2  \alpha A^2 \rho_\chi \bar\sigma_n}{ m_Nm_{\chi}\mu_{\chi n}^2}\int\!\! d^3 v f_\chi(v) \int\!\! d\omega \int\!\frac{d^3\bfq_N}{(2\pi)^3}\int\!\frac{d^3\bfp_f}{(2\pi)^3}\int\!\!\frac{d^3\bfk}{(2\pi)^3}\frac{1}{k^2}\text{Im}\left[\frac{-1}{\epsilon(\bfk,\omega)}\right]   \left[\frac{1}{\omega -\frac{\bfq_N\cdot\bfk}{m_N}}-\frac{1}{\omega}\right]^2\nonumber \\
 &\times |F_{DM}(\bfp_i -\bfp_f)|^2 |F(\bfp_i-\bfp_f-\bfq_N -\bfk)|^2\, \delta\left(E_i-E_f-E_N-\omega\right). 
  \end{align}
\end{widetext}
with $A$, $m_N$ and $\mu_{\chi n}$ are the mass number of the element, the total mass of the nucleus and the DM-nucleon reduced mass, respectively. $\bar \sigma_n$ is the DM-nucleon reference cross section that is used to parametrize the reach. For example, assuming a scalar mediator with universal couplings to protons and neutrons 
\begin{equation}
H=   \ g_\chi \phi \bar\chi\chi + g_n \phi (n\bar n+p\bar p),
\end{equation}
 the reference cross section is defined by
\begin{align}
	\bar \sigma_n = \frac{\mu_{\chi n}^2 g_n^2 g_\chi^2}{\pi \big(q_0^2 + m_{\phi}^2\big)^2}.
\end{align}
where $m_{\phi}$ is the mass of the mediator and $q_0$ is a reference momentum, which we take to be $q_0= m_\chi v_0$ with $v_0$ the DM velocity dispersion. $\Zion$ is the effective charge of the ion, which includes the nucleus and bound core electrons of the atom. $\bfq_N$, $\bfp_f$ and $\bfk$ are the momenta associated with, respectively, the recoiling nucleus, the outgoing DM particle and the electronic excitations. As before, $\omega$ is the energy deposited into electronic excitations, while $E_N\equiv q_N^2/2m_N$ is the kinetic energy of the recoiling nucleus. $F_{DM}(\bfp_i-\bfp_f)$ is the DM-mediator form factor, which was suppressed in \cite{Knapen:2020aky}. It is defined as
\begin{align}\label{eq:DMformfactorNucl}
    F_{DM}(q) \equiv \frac{ q_0^2+m_{\phi}^2 }{q^2+m_{\phi}^2 }.
\end{align}

\begin{figure*}[t]
\includegraphics[width=0.95\textwidth]{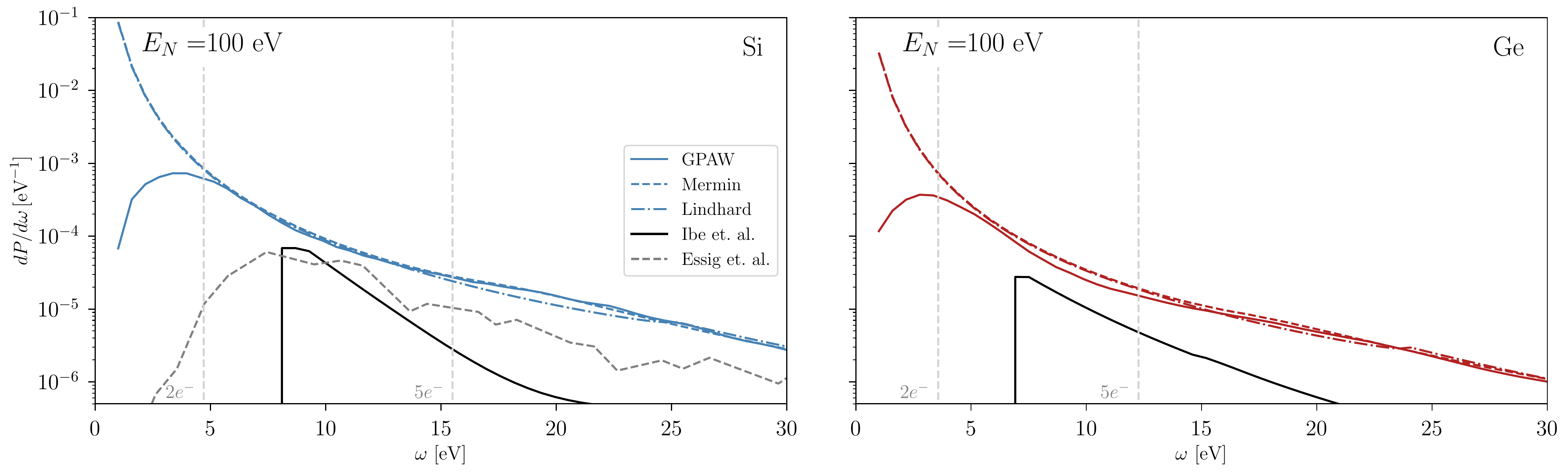}
\caption{Shake-off probability for Si and Ge semiconductors as computed with the Lindhard, Mermin, GPAW methods, and for Si and Ge atomic targets using Ibe~et.~al.~\cite{Ibe:2017yqa}. For comparison, we also show the result in Si semiconductors by Essig~et.~al.~\cite{Essig:2019xkx}. \label{fig:shakeoff}}
\end{figure*}

In a realistic solid, the nucleus is bound to the crystal, which gives rise to the additional crystal form factor $F(\bfp_i-\bfp_f-\bfq_N -\bfk)$ in \eqref{eq:migdalmaster}. In \cite{Knapen:2020aky} we worked in the impulse approximation, which treats the recoiling ion wavefunction as a plane wave, but accounts for the binding potential through the initial state wavefunctions \cite{gunnwarner,watson1996neutron}.  This is valid as long as  $E_N\gg \bar\omega_{ph}$, where $\bar\omega_{ph}$ is the average acoustic phonon frequency in the crystal, of order several tens of meV for most materials. In this limit, the crystal form factor can be approximated by
\begin{equation}
 F(q)\equiv \left(\frac{4\pi}{\Delta^2}\right)^{3/4}e^{\frac{-q^2}{2\Delta^2}}.
 \label{eq:Impulseformfactor}
\end{equation}
with $\Delta\equiv\sqrt{m_N\bar\omega_{ph}}$.
We leave generalizing the calculation beyond the impulse approximation for future work. To avoid extrapolating beyond the regime of validity for the impulse approximation, \code\ will excise the part of phase space for which $E_N$ is below a threshold value $E_{N}^{th}$. By default $E_{N}^{th}$ is taken to be $E_{N}^{th}=4\bar\omega_{ph}$, as explained in Appendix B of \cite{Knapen:2020aky}, but the user can also test different values by setting the \verb+Enth+ parameter. 
Finally, we note that the notation in \eqref{eq:migdalmaster} slightly differs from the notation in \cite{Knapen:2020aky}, since here we suppressed the reciprocal lattice vectors, as explained in Sec.~\ref{sec:ELF}. 

The integral in \eqref{eq:migdalmaster} is rather non-trivial to evaluate due to its high dimension and non-trivial boundary conditions. The problem however simplifies substantially if we approximate the target material as isotropic and work in the soft limit where $|\bfq_N\cdot \bfk|\ll m_N \omega$ and $k\ll q_N$. Estimating $q_N\sim v m_\chi$, the soft approximation holds for \mbox{$10\; \text{MeV} \lesssim m_\chi\lesssim 10$ GeV} and $\omega \gtrsim $ eV, which is the parameter space of interest for the Migdal effect. With these assumptions, \eqref{eq:migdalmaster} can be written as the double differential rate 
\begin{align}\label{eq:migdalfactorized}
\frac{dR}{dE_N d\omega }\approx \frac{\rho_\chi}{ m_Nm_{\chi}}\int\!\! d^3 v\, v f_\chi(v) \frac{d\sigma_{qe}}{d E_N} \frac{d P}{d\omega}
  \end{align}
where we defined the quasi-elastic cross section $\frac{d\sigma_{qe}}{d E_N}$ as
\begin{widetext}
\begin{align}
\frac{d\sigma_{qe}}{d E_N}\equiv\frac{2\pi^2 A^2 \bar\sigma_n}{v \mu_{\chi n}^2} \int\!\frac{d^3\bfq_N}{(2\pi)^3}\int\!\frac{d^3\bfp_f}{(2\pi)^3}
 |F_{DM}(\bfp_i-\bfp_f)|^2 |F(\bfp_i-\bfp_f-\bfq_N)|^2\, \delta\left({E_i-E_f-E_N-\omega}\right)\delta(E_N-\frac{q^2_N}{2m_N})
\end{align}
\end{widetext}
For $\omega= 0$ this quantity reduces to the elastic nuclear recoil cross section. In limit where the nucleus is taken to be a free particle, or $\bar \omega_{ph} \to 0$, the factor $|F(\bfp_i-\bfp_f-\bfq_N)|^2$ moreover asymptotes to $(2\pi)^3\delta(\bfp_i-\bfp_f-\bfq_N)$; then one recovers the familiar result for the elastic recoil of a free nucleus.

The quantity $d P/d\omega$ is the probability density for energy $\omega$ to be deposited into electronic excitations:
\begin{align}
\frac{dP}{d\omega}&= \frac{4\alpha Z_{\rm ion}^2}{\omega^4} \int\!\!\frac{d^3\bfk}{(2\pi)^3} \frac{|\bfv_N\cdot \bfk|^2}{k^2} \text{Im}\left(\frac{-1}{\epsilon(\bfk,\omega) } \right)    \label{eq:Psemi_epsilon}\\
&=\frac{4\alpha Z_{\rm ion}^2E_N}{3\pi^2\omega^4 m_N} \int\!\!d k\,  k^2 \text{Im}\left(\frac{-1}{\epsilon(k,\omega) } \right)
\end{align}
where in the second line we have used the isotropic approximation and $v^2_N=2 E_N/m_N$. The shake-off probability is shown in Fig.~\ref{fig:shakeoff} for Si and Ge, as computed with the Lindhard, Mermin and GPAW methods. Above the 2$e^-$ threshold all three methods are in good agreement, especially for Si. (The reasons for the discrepancy for $\omega\lesssim 5$ eV were explained in Sec.~\ref{sec:DMelectron}.) For comparison, we also show the shake-off probability as obtained using the formalism for the atomic Migdal effect, following~\cite{Ibe:2017yqa}. In this calculation one effectively approximates the system as atomic Si/Ge, neglecting the remainder of the lattice. This approach substantially underestimates the shake-off probability in Si and Ge semiconductors, especially at low $\omega$.

To perform the phase space integrals, we define the following auxiliary functions
\begin{align}
I(\omega)&\equiv \frac{1}{E_N} \frac{d P}{d\omega}\\
J(v,\omega)&\equiv \int\!\!dE_N\,E_N \frac{d\sigma_{qe}}{d E_N}.
\end{align}
where $I(\omega)$ is independent of $E_N$ and $J(v,\omega)$ is the energy-weighted quasi-elastic cross section.
Integrating \eqref{eq:migdalfactorized} over $E_N$, the differential rate in $\omega$ can be written as 
\begin{align}\label{eq:migdaldRdo}
\frac{dR}{d\omega }\approx \frac{\rho_\chi}{ m_Nm_{\chi}}I(\omega) \int\!\! d^3 v\, v f_\chi(v) J(v,\omega).
\end{align}
To speed up the integration, \code\ will tabulate and interpolate $I(\omega)$ with the default settings whenever a \code\ object is initialized. If the flag \verb+fast+ is set to \verb+True+ in the Migdal rate calculations, \code\ will use the precomputed $I(\omega)$ rather than computing it from scratch for each point. The \verb+tabulate_I()+ function can be used to update the precomputed $I(\omega)$ with settings specified by the user. See Appendix~\ref{app:useage} for more details.

\begin{figure*}[t]
\includegraphics[width=0.95\textwidth]{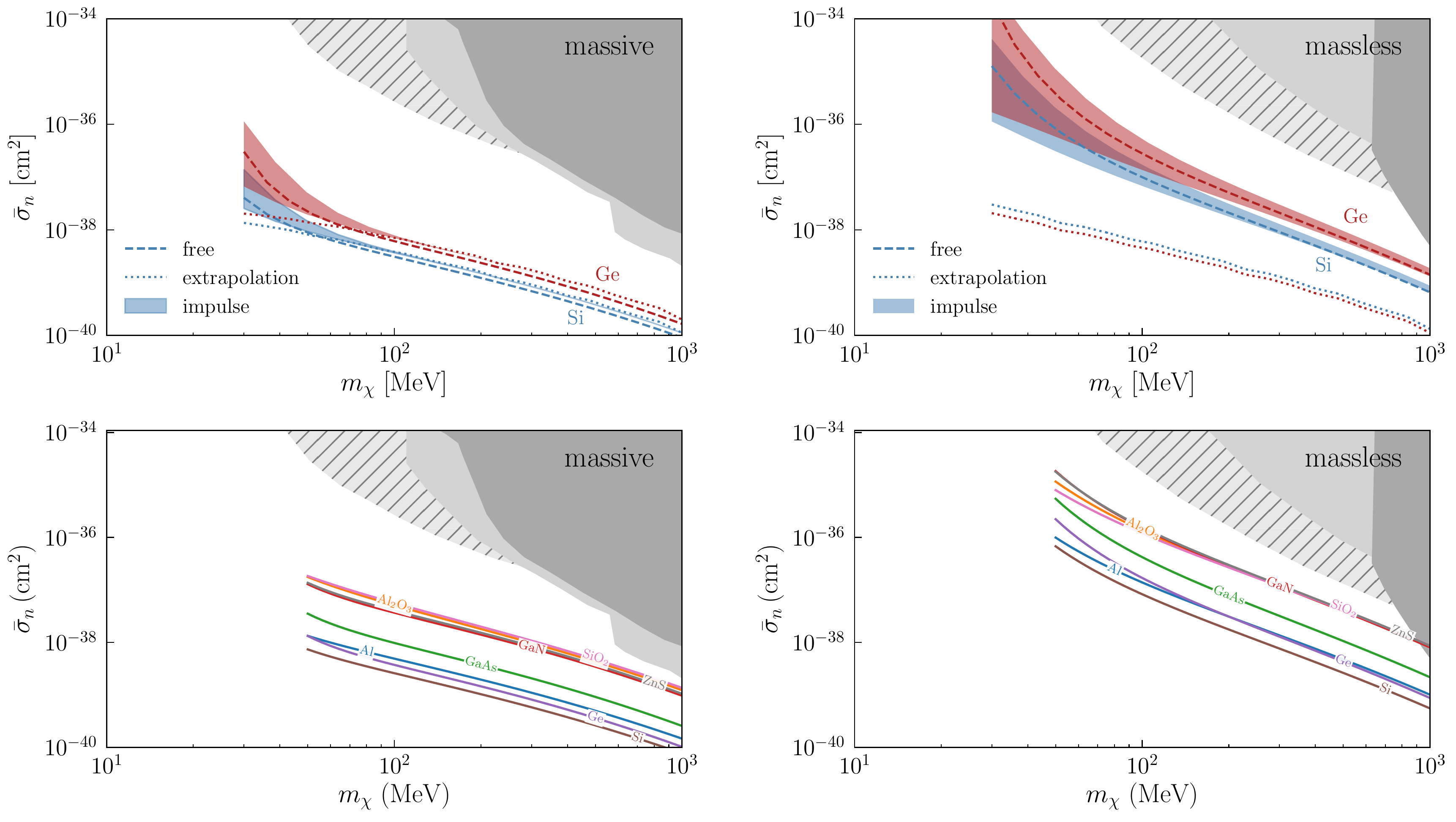}
\caption{\emph{Upper row:} Cross section plots for 3 events with a kg-year exposure, assuming a 2$e^-$ threshold, computed with the GPAW method. We show the massive (left) and massless (right) mediator limits. The dashed lines use the free ion approximation with threshold $E_{N}^{th}=4\bar\omega_{ph}$. The shaded bands use the impulse approximation, varying $E_{N}^{th}$ between $4\bar\omega_{ph}$ and $9\bar\omega_{ph}$, which roughly indicates the uncertainty of the approximation. The dotted lines are an (uncontrolled) extrapolation, where we set $E_{N}^{th}=0$ in the impulse approximation formulas. (See text for details.) The dark gray shaded regions represent nuclear recoil bounds from XENON1T~\cite{Aprile:2019jmx}, LUX~\cite{Akerib:2018hck}, CRESST III~\cite{Abdelhameed:2019hmk} and CDEX~\cite{Liu:2019kzq}, while the light gray region is the recent XENON1T limit using the Migdal effect~\cite{Aprile:2019xxb}. The hashed regions are recasted XENON limits in terms of the Migdal effect by Essig~et.~al.~\cite{Essig:2019xkx}. \emph{Bottom row:} Same as top row, but for a wider range of materials, using the Mermin method and the free ion approximation. For materials with multiple types of atoms, we approximate the rate as coming from the heavier atom. \label{fig:migdal_money}}
\end{figure*}

It thus remains to evaluate $J(v,\omega)$. In the free ion approximation, the crystal form factor squared asymptotes to a delta function, and we obtain the closed-form expression
\begin{align}
J(v,\omega)=&\frac{A^2 \bar\sigma_n}{8 v^2 m_N  \mu_{\chi n}^2}(q_0^2+m_{\phi}^2)^2\Bigg[\log\left(\frac{q_{+}^2+m_{\phi}^2}{q_{-}^2+m_{\phi}^2}\right)\nonumber\\
&+\frac{m_{\phi}^2}{q_{+}^2+m_{\phi}^2}-\frac{m_{\phi}^2}{q_{-}^2+m_{\phi}^2}\Bigg] 
\end{align}
with 
\begin{align}
q_{-}&=\text{max}\left[v\mu_{\chi N}\left(1-\sqrt{1-\frac{2\omega}{v^2 \mu_{\chi N}}}\right),\sqrt{2m_N E_{N}^{th}}\right]\nonumber\\
q_{+}&=v\mu_{\chi N}\left(1+\sqrt{1-\frac{2\omega}{v^2 \mu_{\chi N}}}\right)
\end{align}
with $E_{N}^{th}$ the energy threshold for the nuclear recoil and $\mu_{\chi N}$ the DM-nucleus reduced mass.  Experimentally, $E_{N}^{th}$ can effectively be zero if one is only interested in the ionization signal. However, theoretically, both the free ion and impulse approximations break down for $E_N^{th}\to 0$ and we therefore use a nonzero $E_{N}^{th}$, as discussed below \eqref{eq:Impulseformfactor}.
In the massive ($m_{\phi}\to \infty$) and massless ($m_{\phi}\to 0$) limits, $J(v,\omega)$ reduces to
\begin{align}
J_{\infty}(v,\omega)&=\frac{ A^2 \bar\sigma_n}{16 v^2 m_N \mu_{\chi n}^2}(q_{+}^4-q_{-}^4)\\
J_{0}(v,\omega)&=\frac{ A^2 \bar\sigma_n q_0^4}{4v^2 m_N \mu_{\chi n}^2}\log\left(\frac{q_{+}}{q_{-}}\right)
\end{align}
where the $\infty$ and $0$ subscripts refer to the massive mediator and massless mediator limits, respectively. 

The expression for $J(\omega,v)$ in the impulse approximation is more complicated: 
\begin{align}\label{eq:Jimpulse}
J(v,\omega)=&\frac{A^2 \bar\sigma_n}{16\sqrt{\pi}v^2m_N \mu_{\chi n}^2}\int_{q^-}^{q^+} \!\! dq\, |F_{DM}(q)|^2\nonumber\\
&\times\left(G(q,q_N^+)-G(q,q_N^-)\right)
\end{align}
with
\begin{align}
G(q,q_N)& \equiv2\Delta (q^2-q q_N +q_N^2+\Delta^2)e^{-\frac{(q+q_N)^2}{\Delta^2}}\nonumber\\
&-2\Delta (q^2+q q_N +q_N^2+\Delta^2)e^{-\frac{(q-q_N)^2}{\Delta^2}}\nonumber\\
&+\sqrt{\pi}q(2q^2+3\Delta^2)\text{Erf}\left[\frac{q-q_N}{\Delta},\frac{q+q_N}{\Delta}\right]
\end{align}
where the incomplete error function is defined as 
\begin{equation}
\text{Erf}(x,y)\equiv\frac{2}{\sqrt{\pi}}\int_{x}^{y}\!\! dt\, e^{-t^2}.
\end{equation}
The boundary conditions are given by
\begin{align}
q_{\pm}&=v\mu_{\chi N}\left(1\pm\sqrt{1-\frac{2\omega}{v^2 \mu_{\chi N}}}\right)\nonumber\\
q^+_N&=\sqrt{2 m_N\left(vq -\frac{q^2}{2m_\chi}-\omega\right)}\nonumber\\
q^-_N&=\sqrt{2 m_N E_{N}^{th}}.
\end{align}
The momentum integral in \eqref{eq:Jimpulse} must be evaluated numerically. As a result, the computation for the impulse approximation is substantially slower than for the free ion approximation.

The cross section plots for a rate of 3 events/kg-year are shown in Fig.~\ref{fig:migdal_money}, where we computed the ELF with the GPAW method and assumed a $2e^-$ threshold. The dashed line is the free ion approximation with $E_{N}^{th}=4\bar\omega_{ph}$. The shaded bands represent the impulse approximation, where we varied $E_{N}^{th}$ between $4\bar\omega_{ph}$ and $9\bar\omega_{ph}$ in order to illustrate the sensitivity to the phase space cut on $E_N$. For $m_\chi\lesssim 30$ MeV this sensitivity becomes very severe and we chose to discontinue the curves. This means that the impulse approximation is not valid in most of the phase space for $m_\chi\lesssim30$ MeV,  and the wavefunction of the ion in the crystal must be accounted for in this regime.
In other words, the energy scale of the DM-nucleus collision is now of the same order as the typical energy scale of acoustic excitations in the crystal, and the Migdal effect must be described in terms of multiphonon processes. At low $m_\chi$, the soft limit we assumed, $k \ll q_N$, also breaks down, since $dP/d\omega$ has non-negligible contributions from $k$ up to $O(10)$ keV.
These sources of uncertainty are much more severe for the massless mediator case, as the DM-mediator form factor biases the rate towards lower momentum transfers and $E_N$. Similar considerations likely also apply to the Migdal effect in liquid Xe, which may affect the limits in~\cite{Aprile:2019jmx,Essig:2019xkx}.

For reference, Fig.~\ref{fig:migdal_money} also shows the result in the impulse approximation where we boldly took $E_N^{th}=0$ (dotted lines).  We emphasize that is an uncontrolled extrapolation, which should \emph{not} be used to obtain sensitivity estimates or limits. It is however useful to understand the robustness of our calculations: In particular, for the massive mediator we see that the dotted line merges with the others for $m_\chi\gtrsim 70$ MeV. In this regime, the part of phase space removed with the $E_N$ cut is a negligible contribution to the total rate, and we expect the result to be unchanged even if one generalizes the computation beyond the impulse approximation. The same is not true for the massless mediator, where the rate is much more heavily weighted toward lower $E_N$. In this case, it is necessary to understand the Migdal effect in the multiphonon regime and away from the soft limit to obtain the total rate. Our current calculation can therefore only be used as a conservative estimate for the massless mediator case.

Finally, the lower panels of Fig.~\ref{fig:migdal_money} show the cross section curves in a wider range of materials, where we assume the free ion approximation with $E_{N}^{th}=4\bar\omega_{ph}$. In materials where there are multiple types of atoms, we estimate the rate by calculating the recoil from the heaviest element only, since we assume that the DM-nucleus cross section scales as $A^2$. The lighter element can contribute a comparable amount, so there are $O(1)$ uncertainties in making this approximation. Still, Si and Ge again have the best reach among semiconductors due to the lower  $2e^{-}$ threshold.


\begin{figure*}[t]
\includegraphics[width=0.48\textwidth]{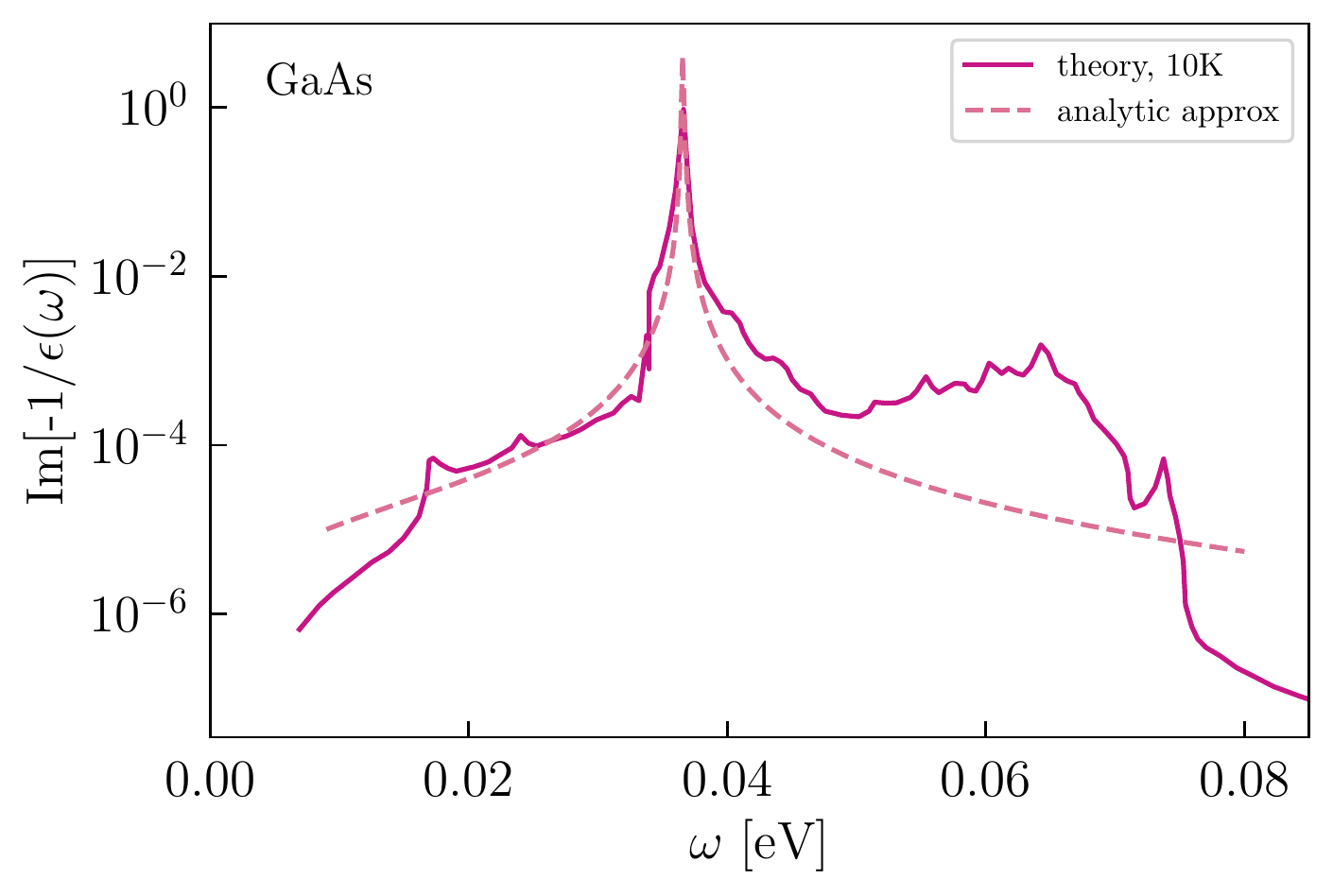}\hfill
\includegraphics[width=0.49\textwidth]{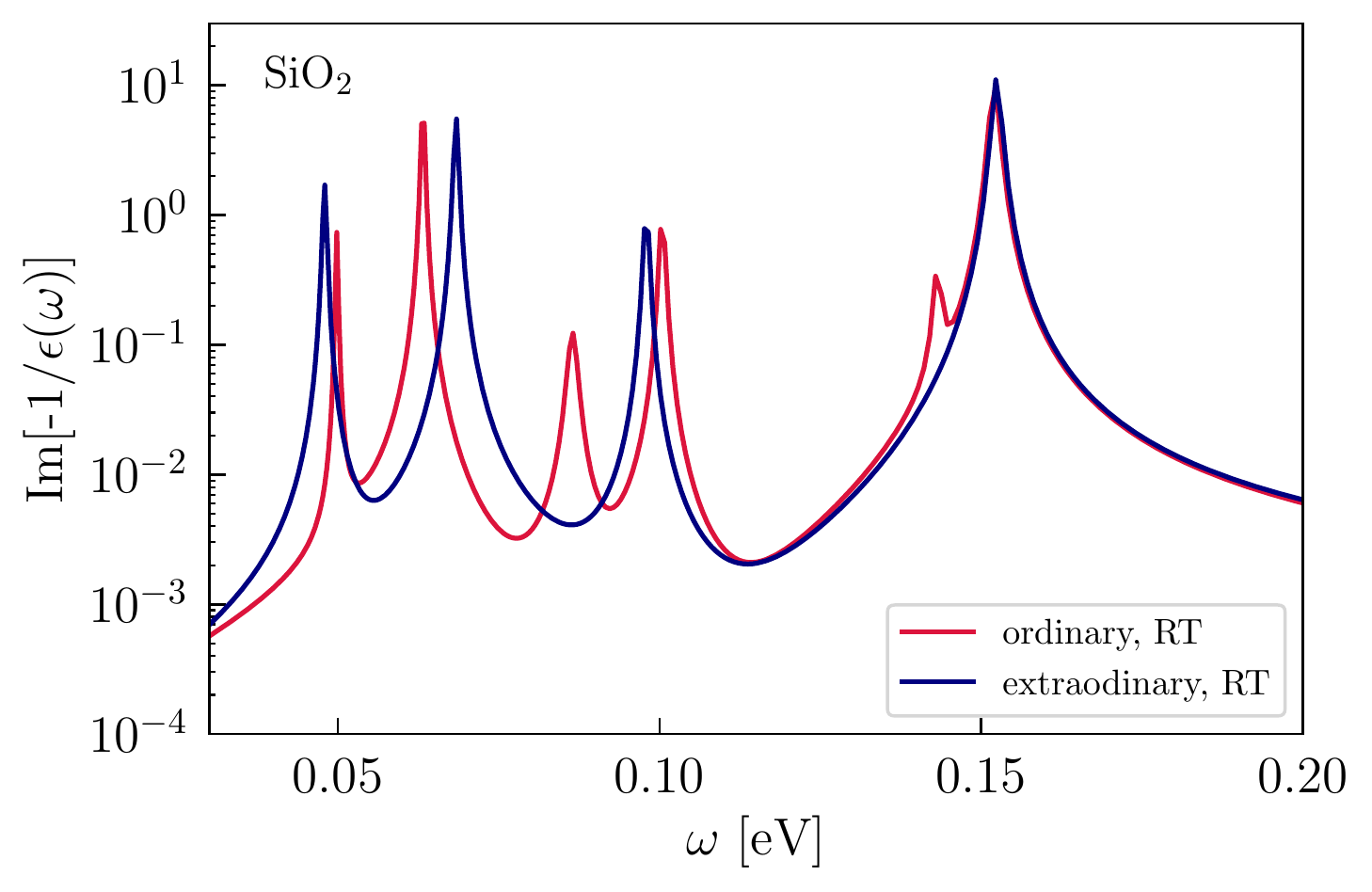}
\caption{Examples of the ELF in the phonon regime and in the optical limit ($k \to 0$). For polar materials, the ELF is dominated by longitudinal optical phonon resonances.  ({\bf Left}) The solid line shows the response obtained from the calculation of absorption at 10 Kelvin~\cite{PhysRevB.70.245209}, combined with the real index of refraction~\cite{opticalconstantsGaAs}. The calculation includes both the optical phonon resonance as well as anharmonic contributions away from the peak. The dashed line shows the response obtained using the analytic approximation of \eqref{eq:epsilon_polarmaterial}, which only partly captures the multiphonon response away from the resonance.
({\bf Right}) We show the response in SiO$_2$ using \eqref{eq:epsilon_polarmaterial} with measured parameters of Ref.~\cite{GervaisSiO2}. The response is shown for ordinary rays ($\vec E \perp c$-axis) and and extraordinary rays ($\vec E \parallel c$-axis) at room temperature. The widths of the resonances depend on temperature and will be smaller at zero temperature; however, since the width drops out in the narrow width limit, this has a negligible impact on the DM-phonon scattering rate. 
\label{fig:phonon_ELF_polar}}
\end{figure*}


\section{Dark matter-phonon scattering \label{sec:DMphonon} }

For energies below the electron band gap, the ELF of a material is dominated by energy loss into phonon excitations. In this section, we discuss how DM-induced phonon excitations can also be treated with the same approach as introduced in \cite{Knapen:2021run} and discussed in Sec.~\ref{sec:DMelectron} above. 

The idea is similar to that of Sec.~\ref{sec:DMelectron}, where now we must consider how the mediator couples to protons, neutrons, and electrons. If the mediator couples to these particles in the same proportions as the SM photon, then we can directly extend the formalism of Sec.~\ref{sec:DMelectron} and apply \eqref{eq:mastereq} below the electron band gap. The intuition behind this result is that an external source can create charge fluctuations in both electrons and ions. The total size of those charge fluctuations determines the dielectric response function $\epsilon^{-1}(\bfk,\omega)$ and thus the energy loss rate. For $\omega$ above the electron band gap, the response is dominated by electrons since the perturbation to the system happens quickly compared the characteristic time-scale of the ion motion in the crystal, $\sim 1/ \bar \omega_{ph}$. For energy deposits \emph{below} the electron band gap, we are in the opposite regime: The response of the electrons is effectively instantaneous on the time scale of the external perturbation. They therefore act as a perfect, dissipationless dielectric. The kinematic degrees of freedom of the ions are now responsible for any energy dissipation in the crystal. 

If the mediator couples to the charge fluctuations differently from the SM photon, then the direct relationship to the dielectric response and ELF will be broken. In the most general case, the dynamic structure factor for phonon excitations must be calculated from first principles according to the mediator couplings. This was discussed in the initial work on this subject~\cite{Knapen:2017ekk,Griffin:2018bjn}, where it was shown for instance that a kinetically-mixed dark photon will lead to optical phonon excitations in polar materials, while a scalar mediator will generally lead to acoustic phonon excitations. The formalism is based closely on the theory of neutron scattering in crystals~\cite{Schober2014}, and further studies of DM-phonon excitations in numerous target materials can be found in Refs.~\cite{Trickle:2019nya,Griffin:2019mvc,Griffin:2020lgd,Coskuner:2021qxo}.

Therefore, in this work we focus on vector mediators which couple to nucleons and electrons in the same way as SM photon. We will work in the massless mediator limit, motivated by cosmological relic benchmarks such as freeze-in~\cite{Essig:2015cda,Dvorkin:2019zdi,Dvorkin:2020xga} in this mass range. The data on the ELF in this regime comes from optical measurements at momentum transfer $k \to 0$, and we will approximate the ELF as being independent of $k$ for this calculation. This is a good approximation for sub-MeV dark matter scattering via ultralight mediators, which is strongly weighted at low momentum transfers $k \ll $ keV, and we show below good agreement with the DFT calculations of \cite{Griffin:2018bjn,Griffin:2020lgd}. With these assumptions, \eqref{eq:mastereq} simplifies to
\begin{align}\label{eq:mastereqphonon}
	R =& \frac{1}{\rho_T} \frac{\rho_\chi}{m_\chi} \frac{  \bar \sigma_e}{\mu_{\chi e}^2} \frac{q_0^4}{4\pi\alpha}\int\! d^3 v\, \frac{f_\chi(v)}{v} \int \frac{d\omega}{2\pi}\nonumber\\
	&\times \Im \left[ \frac{-1}{\epsilon(\omega)} \right] \log\left[\frac{1+\sqrt{1-2\omega/v^2 m_\chi }}{1-\sqrt{1-2\omega/ v^2 m_\chi}}\right]
\end{align}
with the reference momentum $q_0\equiv \alpha m_e$. We also drop the $k$ dependence in the ELF when taking the optical limit.

To make contact with earlier work, we recall that in polar materials, longitudinal optical (LO) phonons generate a long-range polarization in the material, allowing for enhanced interactions with charged particles. For a material such as GaAs with only one LO phonon, the effective coupling of a charged particle with optical phonons is particularly simple and given by the Fr\"{o}hlich Hamiltonian \cite{frolich1954}. The coupling strength of this effective interaction is given by
\begin{align}
    C_F = \sqrt{\frac{\omega_{\rm LO}}{2 } \left(\frac{1}{\epsilon_{\infty}} -  \frac{1}{\epsilon_{0}}  \right)}.
\end{align}
This coupling was discussed in Refs.~\cite{Knapen:2017ekk,Griffin:2018bjn} and applied there to DM scattering into single LO phonons. We now show how to obtain the same Fr\"{o}hlich coupling and DM scattering rate from the ELF, and also generalize it to include multiple optical phonon branches.

To establish the relationship between the ELF and the Fr\"{o}hlich coupling, we use an analytic approximation for the dielectric function in polar materials. This analytic form is also convenient for materials where suitable low temperature ELF data or first principles calculations are not readily available. Concretely, we approximate the dielectric function by~\cite{Gervais}:
\begin{align}
 	\epsilon(\omega) = \epsilon_{\infty} \,  \prod_{\nu} \frac{\omega_{{\rm LO},\nu}^{2} -\omega^{2} - i \omega \gamma_{{\rm LO},\nu}}{\omega_{{\rm TO},\nu}^{2} -\omega^{2} - i \omega \gamma_{{\rm TO},\nu}}
 	\label{eq:epsilon_polarmaterial}
\end{align}
where $\nu$ labels an optical phonon branch containing both longitudinal (LO) and transverse (TO) modes. $\omega_\nu$ and $\gamma_\nu$ are the energy and width of the phonon mode, respectively. $\epsilon_\infty$ is the high-frequency dielectric constant which describe the contribution of electrons to dielectric response below the band gap; that is, this is the dielectric constant at frequencies well above the phonon energies but still below the electron band gap. Using \eqref{eq:epsilon_polarmaterial} allows for excellent fits to optical data along high-symmetry directions of polar crystals, but note that for arbitrary wavevectors the notion of purely transverse and longitudinal optical modes may not be well-defined. In this work, we will mainly work in the isotropic approximation. In general, first-principles approaches to phonon spectra are needed to calculate the full direction-dependent response function, similar to what was done in Refs.~\cite{Griffin:2018bjn,Griffin:2020lgd,Coskuner:2021qxo}

As can be seen from the form of \eqref{eq:epsilon_polarmaterial}, the ELF will be dominated by LO phonon resonances.
Example ELFs for the polar materials GaAs and SiO$_2$ (quartz) are shown in Fig.~\ref{fig:phonon_ELF_polar}.  SiO$_2$ is a birefringent material where the dielectric response depends on the polarization of the incident field with respect to the optical axis (or $c$-axis), with ordinary rays corresponding to $\vec E \perp c$-axis and extraordinary rays corresponding to $\vec E\parallel c$-axis. For transverse photon modes, this therefore corresponds to optical phonon modes with $\bfk \parallel c$-axis (ordinary response) or $\bfk \perp c$-axis (extraordinary response). To determine the response to DM scattering, we must average over the response in different directions for materials which have anisotropic response, which in principle requires determining the full direction-dependent ELF. However, we find in practice that the rate predictions are very similar whether the ordinary or extraordinary response is used. The same conclusion applies to Al$_2$O$_3$ and GaN, which are also birefringent. This is because the rate is usually dominated by a few strong optical phonon modes that do not vary significantly along different directions. For instance, we see that the four strongest modes in the ELF for SiO$_2$ in Fig.~\ref{fig:phonon_ELF_polar} are only shifted slightly between the ordinary and extraordinary response.

To see the connection between the approach here and previous calculations of phonon excitations, note that we can take the narrow phonon width limit since $\gamma_\nu \ll \omega_\nu$ for all materials here. In this limit, we obtain the loss function 
\begin{align}
	\lim_{\gamma \to 0} {\rm Im} \left[ \frac{-1}{\epsilon(\omega)} \right]  = \sum_{\nu}  \pi \delta&\left(\omega - \omega_{{\rm LO},\nu} \right) \frac{\omega_{{\rm LO},\nu}^{2} -\omega_{{\rm TO},\nu}^{2} } {2 \epsilon_{\infty} \, \omega_{{\rm LO},\nu}} \nonumber \\
	&\times 
	\prod_{\mu \neq \nu} \frac{ \omega_{{\rm LO},\nu}^{2} - \omega_{{\rm TO},\mu}^{2} }{ \omega_{{\rm LO},\nu}^{2} - \omega_{{\rm LO},\mu}^{2}  } .
\label{eq:ELF_phonon}
\end{align}
For materials with just a single optical phonon branch, such as GaAs, this simplifies to 
\begin{align}
 \lim_{\gamma \to 0} {\rm Im} \left[ \frac{-1}{\epsilon(\omega)} \right]  & = \pi \delta(\omega -\omega_{\rm LO} ) \times \frac{\omega_{\rm LO}}{2 } \left(\frac{1}{\epsilon_{\infty}} -  \frac{1}{\epsilon_{0}}  \right) \nonumber \\
 & = \pi \delta(\omega -\omega_{\rm LO} ) \times C_F^2 
\end{align}
where in the second line we have identified the Fr\"{o}hlich coupling $C_F$ discussed above. We have also introduced the static dielectric constant $\epsilon_{0} = \epsilon(0) = \epsilon_{\infty}  \omega_{\rm LO}^{2}/\omega_{\rm TO}^{2}$. While $C_F$ as defined here strictly applies only for simple materials with a single optical phonon branch, we can use \eqref{eq:ELF_phonon} more generally given data on the optical phonon frequencies.

\begin{figure}[t]
\includegraphics[width=0.48\textwidth]{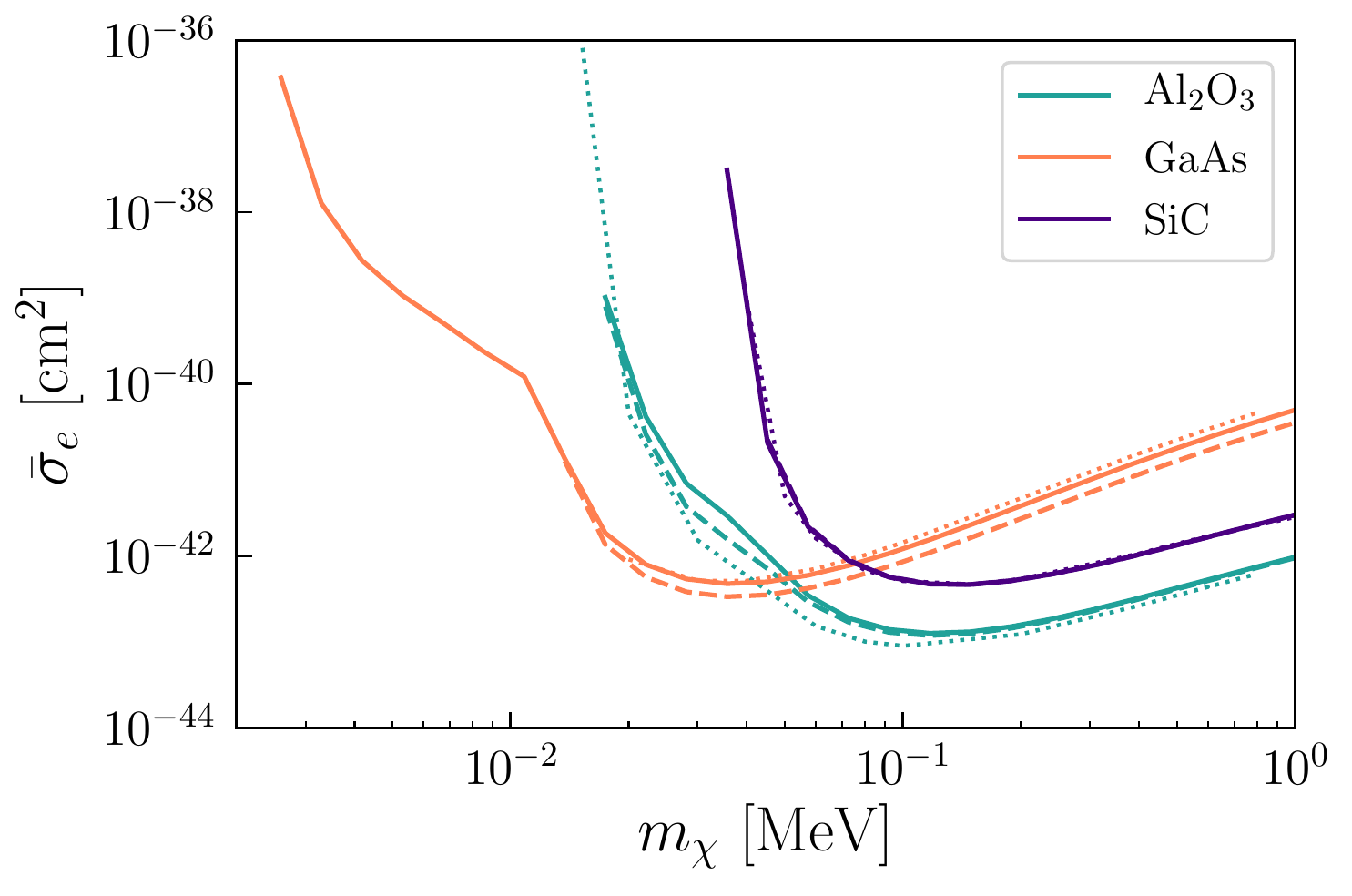}
\caption{Comparison of reach in polar materials, taking different approaches to calculating phonon excitations. The lines shown are the 95\% CL cross section reach with kg-yr exposure and zero background. The result of numerically integrating the ELF over energy (solid lines) agrees well with  the narrow width approximation of \eqref{eq:ELF_phonon} (dashed lines). These further agree well with first-principles numerical calculations of phonon scattering (dotted lines), from Refs.~\cite{Griffin:2018bjn,Griffin:2020lgd}. In the GaAs, the multiphonon response included in the ELF extends the reach to lower masses.
\label{fig:phonon_compare_reach}}
\end{figure}

Fig.~\ref{fig:phonon_compare_reach} compares different approaches to calculating the cross section reach in polar materials. We find good agreement whether we use the full ELF or take the narrow width approximation. (For simplicity, for Al$_2$O$_3$ we use the ordinary dielectric response.) Furthermore, our results line up very well with first-principles numerical calculations of phonon scattering, here taken from Ref.~\cite{Griffin:2018bjn} for GaAs and Al$_2$O$_3$ and from Ref.~\cite{Griffin:2020lgd} for SiC. Note that in the case of GaAs, all approaches agree well for masses above $\sim$10 keV. However, the reach determined by numerically integrating the ELF extends to lower masses, because in this case we use a calculation of the ELF that includes the anharmonic multiphonon response below the optical phonon resonance, as shown in Fig.~\ref{fig:phonon_ELF_polar}. In general, determining the multiphonon response is more challenging, and we only include such contributions where it has been calculated or measured at low temperatures appropriate for a direct detection experiment.

\begin{figure}[t]
\includegraphics[width=0.48\textwidth]{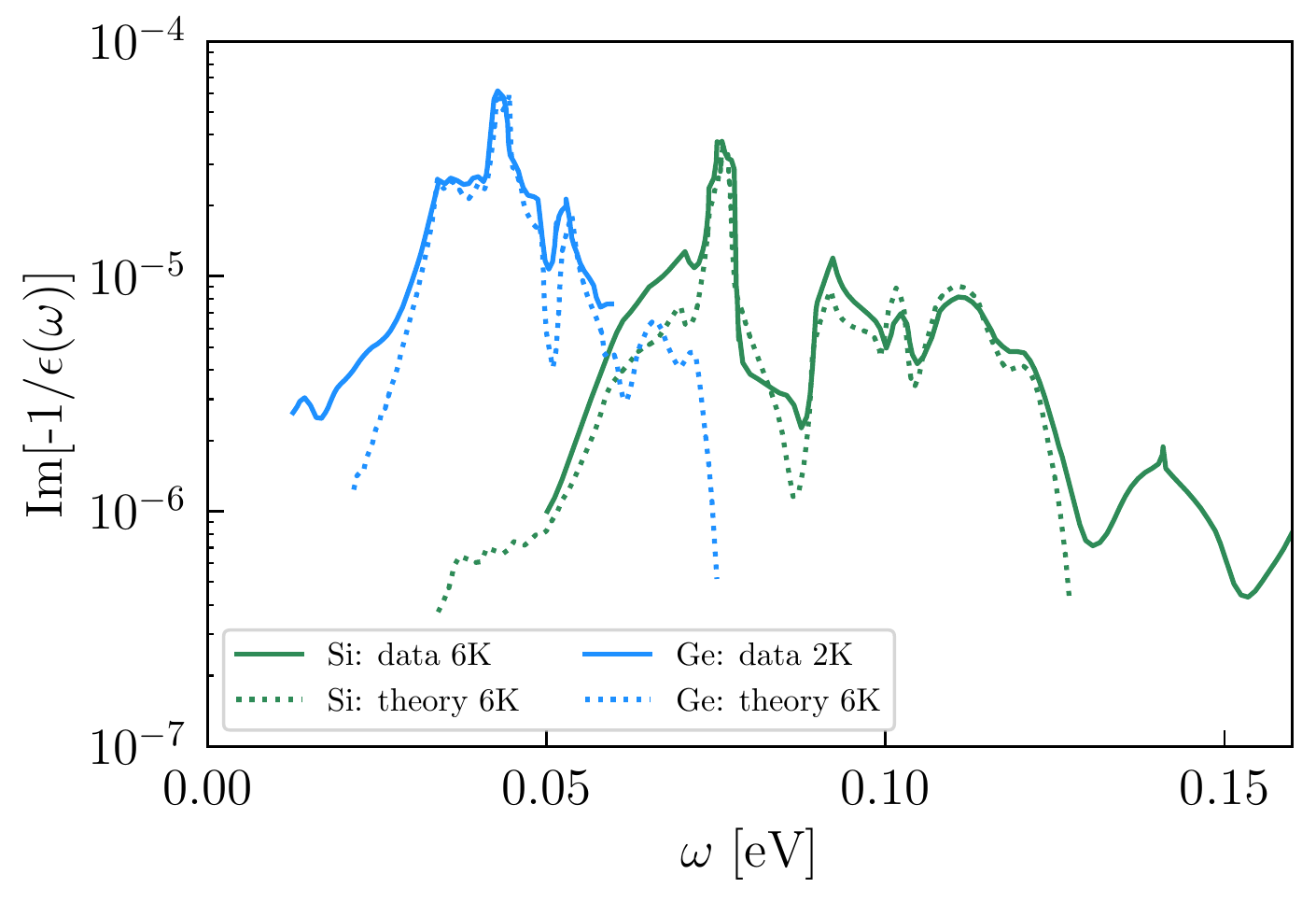}
\caption{For non-polar materials, the optical phonons do not have a long-range polarization and the ELF is instead dominated by multiphonon excitations below the electron band gap. We show the result of optical measurements at 6K for Si~\cite{IkezawaSi} and at 2K for Ge~\cite{IkezawaGe}. The dotted lines show the result of DFT calculations done assuming a temperature of 6K~\cite{PhysRevB.76.054116}.
\label{fig:phonon_ELF_nonpolar}}
\end{figure}

For non-polar crystals, such as Si and Ge, the optical phonon does not have a long-range polarization and the ELF is instead determined entirely by multiphonon excitations. The ELF determined by theory and experiment is shown in Fig.~\ref{fig:phonon_ELF_nonpolar}. Note the overall loss rate is several orders of magnitude smaller than for a polar material.

\begin{table}
\begin{tabular}{|c|c|}
\hline
Material &  ELF in phonon regime \\ \hline
Si & \makecell*{6K data from \cite{IkezawaSi} \\
6K calculation from \cite{PhysRevB.76.054116}}\\
Ge & \makecell*{2K data from \cite{IkezawaGe} \\
6K calculation from \cite{PhysRevB.76.054116}} \\
GaAs  & 10K calculation of \cite{PhysRevB.70.245209}, combined with \cite{opticalconstantsGaAs} \\
Al$_2$O$_3$  & Analytic model, using data from \cite{PhysRevB.61.8187,Gervais} \\
$\alpha-$SiO$_2$ & Analytic model, using 300K data from \cite{GervaisSiO2} \\
GaN  & Analytic model, using 300K data from \cite{GaNdata} \\
ZnS  & Analytic model, using 300K data from \cite{ZnSdata} \\
SiC  & Analytic model of \cite{Griffin:2020lgd}, with data from \cite{Mutschke:1999fg} \\\hline
\end{tabular}
\caption{Sources of the ELF in the phonon regime, for different materials. Analytic model refers to \eqref{eq:epsilon_polarmaterial}, where the references cited have fitted optical data in order to determine the parameters in \eqref{eq:epsilon_polarmaterial} or calculated some of those parameters.  Other cases correspond either to direct measurement or DFT-based calculations of dielectric response. \label{tab:phonon_data}}
\end{table}

\begin{figure}[t]
\includegraphics[width=0.49\textwidth]{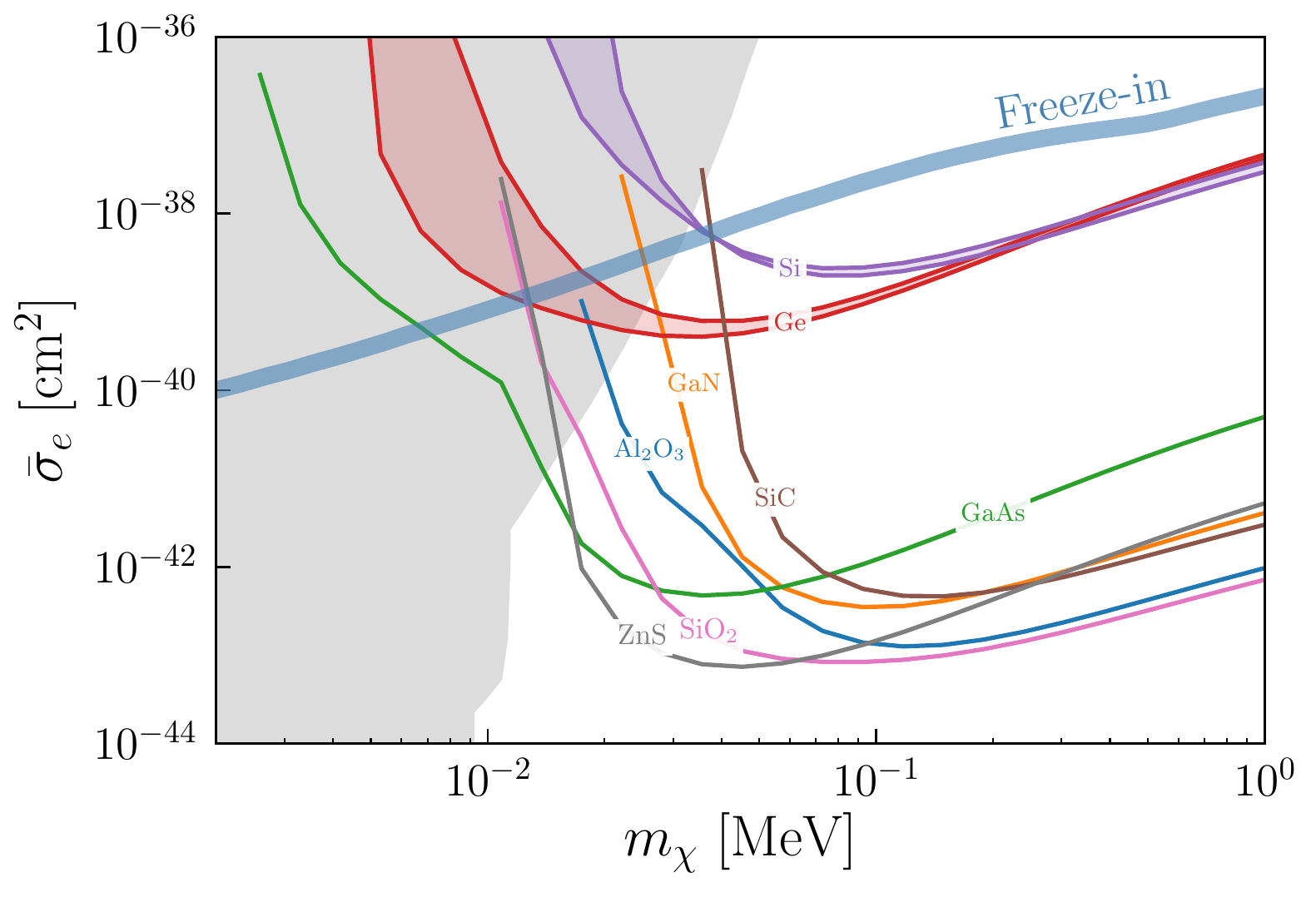}
\caption{Comparison of phonon-based reach from DM scattering in all materials considered here. The lines shown are the 95\% CL cross section reach with kg-yr exposure and zero background.  For Si and Ge, we show both the results obtained using a DFT calculation and using a measurement of the ELF; the region in between is shaded to indicate a rough uncertainty on the true reach. The thick blue line is the predicted cross section if all of the DM was produced by freeze-in~\cite{Essig:2011nj,Dvorkin:2019zdi, Dvorkin:2020xga}. The grey shaded region corresponds to stellar cooling bounds on this DM candidate~\cite{Vogel:2013raa}.
\label{fig:phonon_reach}}
\end{figure}

Fig.~\ref{fig:phonon_reach} summarizes the phonon excitation reach for all materials considered here, and Tab.~\ref{tab:phonon_data} gives the source of the ELF used. Materials like ZnS,  SiO$_2$ and Al$_2$O$_3$ have particularly good reach, due to the fact that they contain strong optical phonon modes down to low energies and because they have a relatively low $\epsilon_\infty$. In particular, $\epsilon_\infty = 5.13$ in ZnS, $\epsilon_\infty = 2.4$ in SiO$_2$ and $\epsilon_\infty = 3.2$ in Al$_2$O$_3$; this  correlates with the higher electron band gap in those materials, thus illustrating the mild tension in optimizing the electron recoil signal vs.~the optical phonon signal in a material. 

Compared to previous studies of the reach in various target materials, the main  advantage here is the inclusion of multiphonon excitations, which are challenging and expensive to compute using first-principles phonon codes. Here we make use of previous studies of multiphonon absorption to determine the multiphonon scattering rate in Si, Ge, and GaAs at low masses. Importantly, the approach described here can give a fast and accurate way to estimate the phonon excitation reach given data or theory on the dielectric response in the phonon regime.

\begin{figure*}[t]
\includegraphics[width=0.99\textwidth]{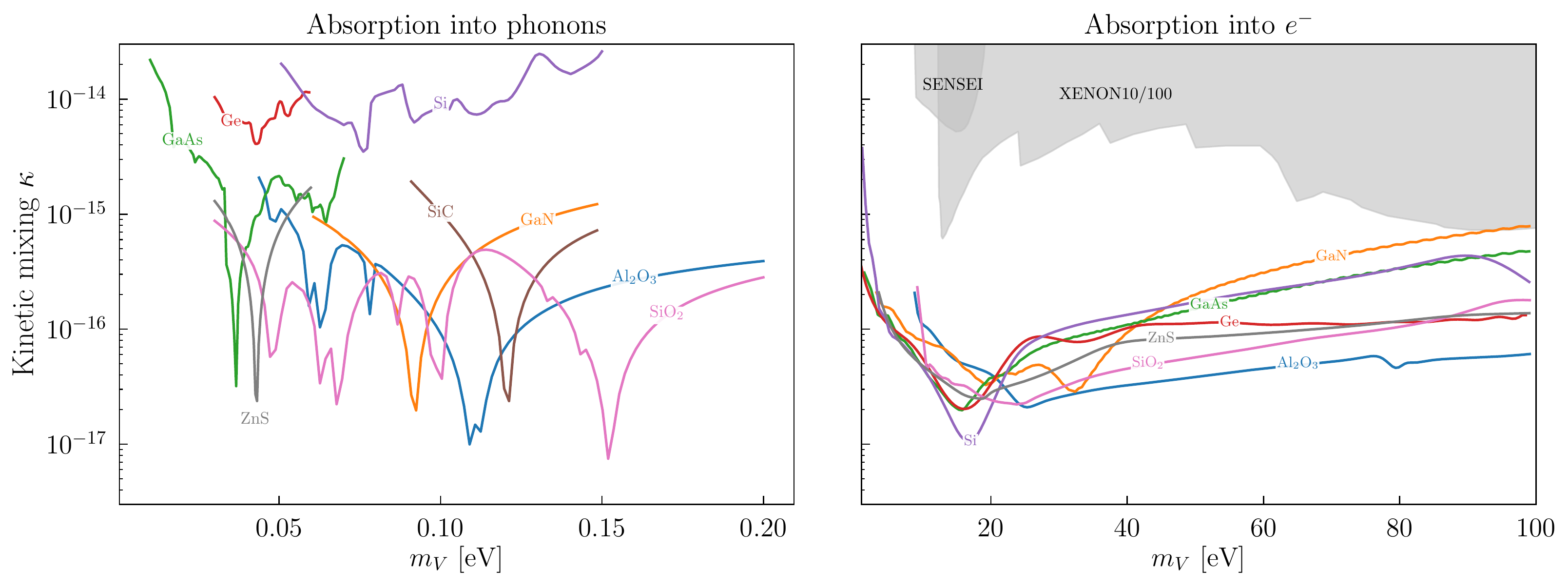}
\caption{A comparison of the reach for absorption of kinetically-mixed dark photon dark matter. The lines shown are the 95\% CL cross section reach with kg-yr exposure and zero background. ({\bf Left}) For phonon excitations, we show here the reach obtained using data on the ELF. As noted in the text, a number of these curves are approximate, given that there is limited data available at zero temperature.  ({\bf Right}) We show here the reach for electron excitations using the Mermin oscillator method for the ELF, and there can be small differences in comparing with DFT methods or direct optical measurements.  The grey shaded regions are limits from XENON10/100~\cite{Bloch:2016sjj} and SENSEI~\cite{Barak:2020fql}. 
\label{fig:absorption_reach}}
\end{figure*}

\section{Absorption of bosonic dark matter \label{sec:absorption}}

Dark matter could also be made up of sub-keV bosons, in which case it can be absorbed by the material into phonon or electron excitations. 
For the specific case where dark matter is comprised of kinetically-mixed dark photons of mass $m_V$, the absorption rate per unit target mass is also determined by the ELF in the zero-momentum limit and given by~\cite{An:2013yua,An:2014twa,Hochberg:2016ajh,Hochberg:2016sqx,Griffin:2018bjn}
\begin{align}
    R = \frac{1}{\rho_T} \frac{\rho_{\rm DM}}{m_V} \kappa^2 m_V \Im \left[ \frac{-1}{\epsilon(m_V)} \right]
\end{align}
where $\kappa$ is the kinetic mixing parameter between the dark and Standard Model photon.
Since optical measurements directly probe the zero momentum limit of the dielectric function, previous works used this data to obtain the absorption rate.

\code\ comes with tabulated ELFs in the optical regime and can therefore be used to quickly obtain the absorption rate. Fig.~\ref{fig:absorption_reach} shows the reach obtained for both phonon and electron excitations for materials included in this work. For electron excitations, the data-driven Mermin method includes optical measurements among the data that is being fitted, and we can take the $k \to 0$ limit of the resulting ELF. Because the Mermin dielectric function does not include an electron band gap, we only use the Mermin ELF for $\omega > E_{\rm gap}$. 

For phonon excitations, we use the same ELF as discussed in Sec.~\ref{sec:DMphonon} and summarized in Tab.~\ref{tab:phonon_data}. Note that some of the data was taken at room temperature and the width of the resonances at sub-Kelvin temperatures will be even smaller. Another caveat to note is that for the birefringent materials (Al$_2$O$_3$, GaN, SiO$_2$) we have taken an average over extraordinary and ordinary response, similar to what was done in Ref.~\cite{Griffin:2018bjn}. This is only approximate and properly accounting for the anisotropy of the material would require a first-principles calculation of the ELF, as discussed in Sec.~\ref{sec:DMphonon}. However, the strongest resonances in the ELF have only a mild direction dependence (see Fig.~\ref{fig:phonon_ELF_polar}), so we expect that the averaging done here gives a good approximation.

\section{Conclusions\label{sec:conclusions}}

We presented \code, a python package to calculate rates for a broad range of DM scattering and absorption processes of interest for direct detection in solid state targets. The unifying feature of these processes is that they are determined by the energy loss function (ELF) of the target material, which characterizes energy loss of Standard Model particles.  \code\ computes  energy loss rates of dark matter particles using tabulated ELFs. At this time, we include ELF data tables for $\mathrm{Al}_2\mathrm{O}_3$, GaN, Al, ZnS, GaAs, $\mathrm{SiO}_2$, Si and Ge assembled from a combination of data, phenomenological models fitted to data, and first-principles calculations.  We aim to add more ELF tables in the future, and our package makes it convenient for users to import their own extractions of the ELF as well.

The currently available dark matter processes, the regime of validity of the calculations, and possible future directions are summarized below:
\begin{itemize}
\item {\emph{ DM-electron scattering}} is determined by the ELF above the electron band gap. We provide ELFs computed in the isotropic limit with a DFT-based method (GPAW) and a data-driven approach (Mermin). Both these approaches start to have large uncertainties at high momentum transfer ($k \gtrsim 20$ keV) which impacts DM-electron scattering at high energies ($\omega \gtrsim 15$ eV) and for scattering via massive mediators. In this regime, improved theoretical calculations and/or data extractions are needed. For instance, to increase the reliability of the Mermin method, a dedicated fit to high $k$ data from a high energy synchrotron facility would be desirable. 
It is also possible to generalize beyond the isotropic approximation and obtain directionally-dependent scattering rates, which would give rise to a daily modulation in strongly anisotropic materials.
\item {\emph{DM-nucleus scattering with Migdal electrons}} depends on the ELF through the probability for a recoiling ion to produce Migdal electrons. The rate to produce Migdal electrons is calculated here for the mass range $30\ \MeV \lesssim m_\chi \lesssim \GeV$. This restriction in mass is due in part to the impulse approximation, which treats the recoiling ion wavefunction as a plane wave. For low nuclear recoil energies that are comparable to typical acoustic phonon energies, a calculation of the Migdal effect with multiphonon production is needed. This will be important if we wish to obtain accurate rates for DM-nucleus scattering via massless mediators and for DM masses below 30 MeV.
\item {\emph{DM-phonon scattering}} is determined by the ELF in the phonon regime, below the electron band gap. Our calculations are valid for DM coupled to a massless kinetically-mixed dark photon mediator, since we use ELF data in the optical limit. While there are already many studies with DFT-based calculations of this process, using existing measurements or calculations of the ELF gives a fast and accurate alternate approach. This approach also incorporates multiphonon contributions, which dominate for non-polar materials and are more challenging to calculate. 
\item {\emph{Absorption of dark photon DM}} has a rate proportional to the ELF in the optical limit $(k =0)$.  Except for the DFT-based calculations in a few cases, the ELFs included are generally obtained either by fitting to optical data or directly from optical data itself. As a result, the ELFs included should describe absorption well in both the phonon and electron regimes.
\end{itemize}

\acknowledgments
We thank Kim Berghaus, Brian Campbell-Deem, So Chigusa, Juan Collar, Rouven Essig, Yonit Hochberg, Yoni Kahn, Robert Lasenby, Noah Kurinsky, Toby Opferkuch,  Diego Redigolo, Tomer Volansky and Tien-Tien Yu for useful discussions. We thank Maarten Vos for providing us with a $\beta$-version of his \texttt{chapidif} package and for his assistance with its usage and the interpretation of the results. TL is supported by the Department of Energy under grant DE-SC0019195 and a UC Hellman fellowship. JK is supported by the Department of Energy under grants DE-SC0019195 and DE-SC0009919.

\appendix

\section{Using darkELF\label{app:useage}}
Here we briefly describe how to run a calculation with \code; for details and examples we refer to the github page.

\paragraph{Conventions:} Natural particle physics units, with $c=\hbar=1$. All masses, momenta and energies are in units of eV. Cross sections are to be specified in units of $\text{cm}^2$.

\paragraph{Dependencies:} \emph{DarkELF} requires python 3.6 or higher, equipped with the \verb+numpy+~\cite{harris2020array}, \verb+scipy+~\cite{2020SciPy-NMeth}, \verb+pyyaml+~\cite{pyyaml} and \verb+pandas+~\cite{mckinney-proc-scipy-2010} packages. The tutorial notebooks require a \verb+jupyter+~\cite{Kluyver:2016aa} installation, but this is in general not needed for \code\ itself.

\paragraph{Setting up calculation:} First be sure that \code\ directory is in your python path. To set up a calculation, the user must first load the package
\begin{verbatim}
from darkelf import darkelf
\end{verbatim}
and subsequently create a darkelf object, which represents a specific target material. This is done by calling the constructor, e.g.~
\begin{verbatim}
Si=darkelf(mX=1e8,mMed=0.0,target='Si',
filename='Si_mermin.dat',
phonon_filename='Si_epsphonon_data6K.dat')
\end{verbatim}
where the \verb+filename+ refers to the precomputed look-up table for the dielectric function in the electronic regime, and \verb+phonon_filename+ sets the look-up table for the phonon regime. The former is tabulated as a function of both $\omega$ and $k$, while the latter only in terms of $\omega$, assuming $k=0$. For some materials, \code\ provides multiple look-up tables for the same ELF, but obtained with different methods, e.g.~Mermin vs.~GPAW. The \verb+filename+ and \verb+phonon_filename+ flags allow the user to specify the ELF computation of their choice. It is possible to leave \verb+phonon_filename+ and \verb+filename+ unspecified, however in this case the various functions relying on the omitted look-up table will be unavailable. For example, users only interested in $e^-$ recoils or the Migdal effect can leave the \verb+phonon_filename+ flag unspecified but must specify \verb+filename+. The dark matter mass and mediator type will also be set at this stage, respectively with the \verb+mX+ and \verb+mMed+ flags. If they are left unspecified, \code\ will set them to the default values. The user must create a separate \verb+darkelf+ object for each target material under consideration. 

The dark matter and mediator masses stored in the \verb+darkelf+ object can be updated by running the \verb+update_params+ method, for example
\begin{verbatim}
Si.update_params(mX=1e7,mMed=1e6)
\end{verbatim}
sets the DM and mediator masses to 10 MeV and 1 MeV respectively. As an alternative to setting the mediator mass with the \verb+mMed+ flag, the \verb+mediator='massless'+ or \verb+mediator='massive'+ flags can be used to specify the massless and massive mediator limits respectively. 

The real and imaginary parts of the dielectric function and the ELF can be accessed by running
\begin{verbatim}
Si.eps1(om,k,method='grid')
Si.eps2(om,k,method='grid')
Si.elf(om,k,method='grid')
\end{verbatim}
with \verb+om+ and \verb+k+ the energy and momentum, both in units of eV. The \verb+method+ flag can take values \verb+grid+, \verb+Lindhard+ or \verb+phonon+, with \verb+grid+ being the default. If the method is \verb+grid+, then $\epsilon_{1,2}$ are obtained from an interpolation of the grid supplied in the \verb+filename+ flag, \verb+Si_mermin.dat+ in the example above. (As the filename indicates, this particular grid was computed with the Mermin method.) This grid applies to the electronic ELF and can be a pre-computed grid with the Mermin or GPAW method, or a grid supplied by the user. The \verb+Lindhard+ flag invokes the Lindhard model in \eqref{eq:eps_lindhard_gas}, which only relies on the plasma frequency. The latter is set in the \verb+.yaml+ file associated with the target material. Finally, if the \verb+phonon+ flag is set, \code\ will use the phonon ELF, which must be set with the \verb+phonon_filename+ flag in the object constructor. The phonon ELF is always computed or measured in the optical limit, and the momentum parameter \verb+k+ is therefore ignored for the \verb+method=phonon+ setting.

\paragraph{Electron recoils:} \code\ can compute the overall rate and differential distributions for DM-electron recoils, with the functions listed in Tab.~\ref{tab:codefunctionsER}. The rate functions allow for optional arguments \verb+sigmae+, \verb+withscreening+ and \verb+method+. \verb+sigmae+ allows the user to change the reference cross section, which is by default set to $\bar\sigma_e=10^{-38}\text{cm}^2$. The boolean flag \verb+withscreening+ enables the user to turn off screening effects, to facilitate comparison with earlier results in the literature. The default value is  \verb+withscreening=True+. Finally, the \verb+method+ flag allows the user to specify the method used for computing the ELF, which must be either \verb+grid+ or \verb+Lindhard+ (see above). 

In addition, \verb+dRdomega_electron(omega)+, \verb+dRdQ_electron(Q)+ and \verb+R_electron()+ have the optional argument \verb+kcut+, which specifies the upper bound on the momentum $k$ that is included in the phase space integral. By default, \code\ will use the kinematical boundary condition or the endpoint of the ELF grid to cut off the $k$ integration, whichever is lower.  \verb+kcut+ allows the user to overwrite this behavior, which can be useful if one is interested in comparing rates for the low momentum part of phase space only. Finally, \verb+R_electron()+ has the optional flag \verb+threshold+, which specifies the lower threshold when integrating over $\omega$. By default, this value is the two $e^-$ threshold for Si, Ge and GaAs and twice the band gap for the remaining materials. \code\ also has a few small auxiliary functions which converts the energy $\omega$ to the number of ionization electrons following \cite{Essig:2015cda}, and the a method to convert the effective millicharge of the dark matter to the reference cross section $\bar\sigma_e$, in the massless mediator limit.

\begin{table*}
\begin{tabular}{p{0.27\textwidth}p{0.50\textwidth}p{0.2\textwidth}}
\multicolumn{3}{c}{electron recoils}\\\hline
function&description& available for\\\hline
\verb+dRdomegadk_electron(omega,k)+&$d^{2}R/d\omega dk$: counts / kg-year $\times$ eV$^2$&all except Xe, SiC and C\\
\verb+dRdomega_electron(omega)+&$dR/d\omega$:  counts / (kg-year $\times$ eV)&all except Xe, SiC and C\\
\verb+dRdQ_electron(Q)+&$dR/dQ$:  counts / (kg-year), binned in \# ionization $e^-$ &Si, Ge, GaAs\\
\verb+R_electron()+&$R$:  counts / kg-year&all except Xe, SiC and C\\
\verb+electron_yield(omega)+&Converts energy to number of ionization electrons&Si, Ge, GaAs\\
\verb+sigmaebar(Qx,mX)+&$\bar \sigma_e$ in terms of $m_\chi$ and $Q_\chi$  for massless dark photon mediator&all except Xe, SiC and C\\\hline
\end{tabular}
\caption{List of public functions in \code\ that relate to electron recoils. Only mandatory arguments are shown; for optional arguments and flags we refer to the text and the documentation in repository. Some functions are only available for select materials, as indicated in the righthand column. Here $Q$ indicates the number of ionization electrons and $Q_\chi$ the effective milicharge of the DM, in the massless mediator limit.\label{tab:codefunctionsER}}
\end{table*}

\paragraph{Migdal effect:} \code\ can compute the shake-off probability as well as the overall and differential rate for the Migdal effect. The public functions related to the Migdal effect are listed in Tab.~\ref{tab:migdal}. The \verb+dPdomegadk+, \verb+dPdomega+ and \verb+tabulate_I+ functions have the optional arguments \verb+method+, \verb+kcut+ and \verb+Nshell+. \verb+method+ can take on the values \verb+Lindhard+, \verb+grid+ and \verb+Ibe+. The former two method, as well as the \verb+kcut+ flag, work as described above. The \verb+Ibe+ option returns the shake-off probability computed using the atomic wave functions in Ibe~et.~al.~\cite{Ibe:2017yqa}. The \verb+Nshell+ flag must be set to an integer and denotes the number of shells included in the atomic calculation. It is ignored if \verb+method+ is set to \verb+Lindhard+ or \verb+grid+. 

In addition, \verb+dRdomega_migdal+ also takes the optional arguments \verb+Enth+, \verb+sigma_n+, \verb+approximation+ and \verb+fast+. \verb+Enth+ corresponds to threshold nuclear recoil energy $E_N^{th}$ and \verb+sigma_n+ is the reference DM-nucleon cross section $\bar\sigma_n$. The \verb+approximation+ flag can be set to \verb+free+ or \verb+impulse+, to toggle between the free ion and impulse approximations. The latter is more accurate though, the former is substantially faster.  The \verb+fast+ flag is a Boolean which specifies whether or not the pre-tabulated values for the shake-off probability are used. Setting \verb+fast=True+ speeds up the calculation but can be inconvenient if one desires to compare different settings for the shake-off probability for a small number of example points. Finally, \verb+R_migdal+ takes the same arguments as \verb+dRdomega_migdal+, in addition to \verb+threshold+, which sets the energy threshold for the electronic excitations.  Note that currently the Migdal calculation in \code\ only accounts for the heaviest element in multi-atomic materials such as $\mathrm{Al}_2\mathrm{O}_3$ and GaAs, assuming that it dominates when the DM-nucleus cross section scales as $A^2$. Generalizing this to include all elements in the crystal is left for future developments.

\begin{table*}
\begin{tabular}{p{0.23\textwidth}p{0.6\textwidth}p{0.15\textwidth}}
\multicolumn{3}{c}{Migdal effect}\\\hline
function&description& available for\\\hline
\verb+dPdomegadk(omega,k,En)+&$d^{2}P/d\omega dk$: shake-off probability, in units of 1/eV$^2$&all except SiC\\
\verb+dPdomega(omega,En)+&$dP/d\omega$: shake-off probability, in units of 1/eV&all except SiC\\
\verb+tabulate_I()+&Tabulates shake-off probability for faster computations&all except SiC\\
\verb+dRdEn_nuclear(En)+&$dR/dE_N$ for elastic nuclear recoils, in units of counts / (kg-year $\times$ eV)&all\\
\verb+dRdomega_migdal(omega)+&$dR/d\omega$ for Migdal effect,  in units of counts / (kg-year $\times$ eV)&all except SiC\\
\verb+R_migdal()+&$R$ for Migdal effect, in units of counts / kg-year&all except SiC\\\hline
\end{tabular}
\cprotect\caption{List of public functions in \code\ that relate to the Migdal effect. Only mandatory arguments are shown; for optional arguments and flags, see text and the documentation in the repository. Some functions are only available for select materials, as indicated in the righthand column. The \verb+Ibe+ option only is available for Si, Ge, C and Xe. For C and Xe the \verb+grid+ option is unavailable.  \label{tab:migdal}}
\end{table*}

\paragraph{DM-phonon scattering:} The double differential, differential and total DM-phonon scattering rate is computed with the functions \verb+dRdomegadk_phonon+, \verb+dRdomega_phonon+ and \verb+R_phonon+ respectively. All three routines accept the optional flag \verb+sigmae+, which sets the effective electron cross section defined in \eqref{eq:sigmaedefinition}, in units of $\text{cm}^2$. The \verb+R_phonon_Frohlich+ function is the same as \verb+R_phonon+ but uses the Fr\"{o}hlich analytic approximation instead of the ELF method. Note that these rates should only be applied for the massless mediator limit since data at large $k$ is not included for the ELF in the phonon regime.

\begin{table*}
\begin{tabular}{p{0.25\textwidth}p{0.5\textwidth}p{0.22\textwidth}}
\multicolumn{3}{c}{DM-phonon scattering}\\\hline
function&description& available for\\\hline
\verb+dRdomegadk_phonon(omega,k)+&Double differential phonon  rate $dR/d\omega dk$ in 1/kg/yr/eV$^2$&all except Al, C and Xe \\
\verb+dRdomega_phonon(omega)+&Differential phonon  rate $dR/d\omega$ in 1/kg/yr/eV&all except Al, C and Xe\\
\verb+R_phonon()+&Total phonon  rate in 1/kg/yr&all, except Al, C and Xe\\
\verb+R_phonon_Frohlich()+&Total phonon rate in 1/kg/yr with analytic approximation&all except Al, C and Xe\\\hline
\end{tabular}
\caption{List of public functions in \code\ related to DM-phonon scattering. Only mandatory arguments are shown; for optional arguments and flags, see text and the documentation in repository. Some functions are only available for select materials, as indicated in the righthand column. \label{tab:phonon}}
\end{table*}

\paragraph{Dark photon absorption:} \code\ can compute the absorption rate for dark photon DM into both phonons and electronic excitations. The computation can be accessed through the \verb+R_absorption+ routine (see Tab.~\ref{tab:absorption}), which has one optional parameter \verb+kappa+, which sets the mixing parameter between the dark photon and the SM photon. \code\ uses the dark matter mass to automatically determine whether the phonon or electron ELF must be used. \verb+R_absorption+ returns 0 if $m_\chi$ is outside the range of the available ELF grids.

\begin{table*}
\begin{tabular}{p{0.24\textwidth}p{0.55\textwidth}p{0.18\textwidth}}
\multicolumn{3}{c}{Absorption}\\\hline
function&description& available for\\\hline
\verb+R_absorption()+&rate for dark photon absorption, in units of counts / kg-year&all except C and Xe\\\hline
\end{tabular}
\caption{List of public functions in \code\ that relate to absorption processes. Only mandatory arguments are shown; for optional arguments and flags, see text and the documentation in the repository. Some functions are only available for select materials, as indicated in the righthand column.  \label{tab:absorption}}
\end{table*}

\paragraph{Adding new materials and/or look-up tables:}
To add a new ELF look-up table, simply add the file to the data folder of the relevant material and load the new grid in the constructor of the darkelf object with the \verb+filename+ or \verb+phonon_filename+ flag, as described above. For ELF in the electronic regime, the data format of the look-up table should be a 4 column, tab separated text file, where the columns represent $\omega$, $k$, $\epsilon_1$ and $\epsilon_2$, with $\omega$ and $k$ in units of eV. For ELFs in the phonon regime, the format is instead $\omega$, $\epsilon_1$ and $\epsilon_2$. To add a new target material, first  create a new subfolder in the data folder named after the material of interest. Then add a~.yaml file to the new folder in which one should specify the various global properties of the material, such as the plasma frequency, mass density etc. The name of the the~.yaml file must match the name of the folder. Any precomputed ELF look-up tables also go in this folder. Finally, the material can be loaded by the setting \verb+target+ flag in the \verb+darkelf+ constructor to the name of the folder corresponding to the new material. 

\bibliography{dielectric}

\end{document}